\documentclass[useAMS,usenatbib]{mn2e}
\onecolumn
\usepackage{epsfig} 
\usepackage{amsmath}
\title[Matching Catalogues by Probabilistic Pattern Classification]{Matching of Catalogues
  by Probabilistic Pattern Classification}
\author[D. J. Rohde et al]{D. J. Rohde$^{1,2}$\thanks{E-mail:
djr@physics.uq.edu.au},  M. R. Gallagher$^{2}$, M. J. Drinkwater$^{1},$
  K. A. Pimbblet$^{1}$\\
$^{1}$Department of Physics, University of Queensland, Queensland, 4072, Australia\\
$^{2}$School of ITEE, University of Queensland, Queensland, 4072, Australia\\
}

\begin{document}

\date{}

\pagerange{\pageref{firstpage}--\pageref{lastpage}} \pubyear{2002}

\maketitle

\label{firstpage}
\begin{abstract}

We consider the statistical problem of catalogue matching from a machine
learning perspective with the goal of producing probabilistic outputs, and
using all available information.  A framework is provided that unifies two
existing
approaches to producing probabilistic outputs in the literature, one based on
combining distribution
estimates and the other based on combining probabilistic classifiers.  We
apply both of these to the problem of matching the HIPASS radio catalogue
with large positional uncertainties to the much denser SuperCOSMOS
catalogue with much smaller positional uncertainties.  We demonstrate the
utility of probabilistic outputs by a controllable completeness and
efficiency trade-off and by identifying objects that have high probability
of being rare.  Finally, possible biasing effects in the
output of these classifiers are also highlighted and discussed.

\end{abstract}


\section{Introduction}
The Virtual Observatory (VO) aims to enable new science  
by enhanced access to data and, more importantly, providing the 
computing resources required to analyze the data (see www.aus-vo.org 
for the Australian contribution to the VO).  One of the most
important capabilities of the VO will be the identification of
different observations of the same object.
A promising VO tool developed for this task is the Open SkyQuery 
protocol \citep{2004adass..13..177B}.  This tool encourages the combination 
of many disparate catalogues and will, in the long term,
offer a powerful aid to VO enabled science.

A problem presents itself, however, when
attempting to combine catalogues with significantly different positional
resolutions.  A salient example of this is the study of \citet{doyle}
who matched the HI Parkes All Sky Survey (HIPASS) (radio) catalogue
\citep{meyer} to the SuperCOSMOS (optical photographic survey)
catalogue \citep{hambly}.
The HIPASS catalogue is significantly less dense in terms of objects
per unit solid angle and has larger positional uncertainties
than the SuperCOSMOS catalogue, which by contrast, possesses much
more accurate astrometry.
This leads directly to objects in the HIPASS catalogue 
having multiple candidate counterparts to the objects in the 
SuperCOSMOS catalogue (Fig~\ref{mismatch1}).  Previously \citep{rohde2}
we borrowed the term linkage from the computer science term of record
linkage in order to emphasise the statistical aspect of the problem of
finding different observations of the same object.  In the literature
the terms matching, associations and cross-referencing are also used
to refer to this (statistical) problem.  Here we use the terms matching and
linkage interchangeably.

The methods we consider here would be classed as empirical Bayes.
Empirical Bayes is a method where frequentist estimators are made of
underlying distributions.  These estimators are then treated as if they 
are completely true.  We then use theory of   
probability in order to calculate class membership (match or non-match).  
The probability that a candidate matches the sparse object 
is conditioned on (informed by) available information, such as its position, flux
and other measurements as well as the observed parameters of other
candidates.  It is the goal of our work to condition on \emph{all} available information.

Two distinct approaches to catalogue matching or linkage have emerged in the literature, however each
fails to use all available information.  The first is
the generative \footnote{Here the term generative refers to
the fact that our model consists of distributions that can `generate'
more data.}  approach of \cite{sutherland}  where a number of probability 
density functions associated with each candidate are combined to
provide an overall probability.  This method is typically applied only in low dimensions of
one or two parameters which means that some potentially useful
information is ignored.  The second approach is a
discriminative pattern classification approach that
the author has taken in \citet{rohde,rohde2} and has also been explored in
\citet{voisin}.  The outputs of these discriminative pattern classification algorithms are
binary, that is, they do not 
provide any indication of the confidence of class membership (match or non-match),
however simple extensions such as those outlined in \citet{platt}
allow for the output to be converted into a probability.
We use the term discriminative to refer to a classifier that gives a probability of class
membership.  There are arguments that each approach has advantages and
disadvantages, we discuss some of these latter.

The discriminative pattern classification approach seems well suited to dealing with
high dimensional distributions; however it has a significant drawback
in that it calculates the probability that an object matches
conditioned on the parameters for that object only.  This could lead to
inconsistencies such as the sum of all candidate probabilities not
summing to one.  What is
required is a probability that an object is a match conditioned
on all available information (i.e. the parameters of all candidate
objects and the parameters of the object being matched to).  A major
result of this paper is the formulation of a method to combine these
probabilities in ignorance of candidate information to a probability
conditioned on all available information. 

From our point of view there are a number of benefits in building a model that
produces probabilistic outputs.  Firstly our formalism relies on
intermediate probabilities to be calculated in order to combine all
available information.  Secondly the probabilistic output allows difficult matches to be
discarded from a scientific analysis.  Thirdly we demonstrate how it
is possible to use probabilities in order to assist in the search for
rare objects.  We demonstrate this by finding dark galaxy candidates
i.e. HI sources from the HIPASS catalogue with a relatively high
probability of having no candidate match.  Generative models are
inherently probabilistic.  Discriminative models can either represent
decision (or classification) boundaries or can give probabilities of class membership.

The importance of the distinction between the two techniques is that
it is often suggested that it is easier or better to estimate
$P(C=1|x)$, where $C=1$ refers to class is one (of two) and $x$ is a 
high dimensional input vector\footnote{Typically $x$ includes all
parameters that might aid in classification in our application.  It
would include position, magnitude, area, colour, flux and redshift.} 
rather than estimating component 
probabilities $P(x|C=1)$, $P(x|C=2)$ and $P(C=1)=1-P(C=2)$ and applying
Bayes' rule.  This argument is particularly common when $x$ is high
dimensional.  We make a comparison of methods applying both and consider
the arguments for and against each formulation.  We conclude that our
final results are largely indifferent to the choice of formulation.
Finally we make a case that probabilities are useful in that they 
allow a trade-off between completeness and efficiency to be achieved.

We further demonstrate
the use of probabilities by considering the 
problem of identifying dark galaxy candidates.  In \citet{doyle} a
search for HI objects with no optical detection (dark galaxies) was
conducted.  No strong candidate for objects with this property was
found.  All HI objects were either accompanied by optical galaxies, or
the absence of optical galaxies was satisfactorily explained by the
field being obscured by dust or stars.  As such the result of Doyle's study
was to conclude that HIPASS did not detect any isolated dark
galaxies.  

If a dark galaxy were to be detected by HIPASS there is high
probability that it could not be identified as such because unrelated background
optical galaxies fall within HIPASS's large positional uncertainty.
The probabilistic output of our classifier is ideal for identifying HI
objects with high probability of having no match.  We present a list
of objects that would be interesting targets for follow up observation.

In Section~\ref{theproblem}, 
we introduce and review the problem in detail, drawing a
connection between the \cite{sutherland} approach and machine learning methods.  
Section~\ref{algorprinc} details the
algorithms for probabilistic classification.
In Section~\ref{applic} we apply our methods to 
the HIPASS-SuperCOSMOS problem
and evaluate the usefulness of our technique.  
We discuss biasing limitations on the application of this method to
scientific problems in Section~\ref{future}.

\section{The Problem}
\label{theproblem}

Our previous work \citep{rohde} developed a classifier that predicted
if a candidate was a match or a non-match conditioned on the
parameters of the candidate in question only.  The \cite{sutherland} formalism,
however, 
conditions the probability on the
parameters of \emph{all} candidates.  In this section we
outline how it is possible to understand both approaches in the same
probabilistic framework.  The formulation of the problem presented
here is influenced by \citet{fellegi} who develop similar ideas for
textual data, using the term record linkage.

\subsection{Framework for Matching}

\begin{figure}
\center
\includegraphics[width=3.5in]{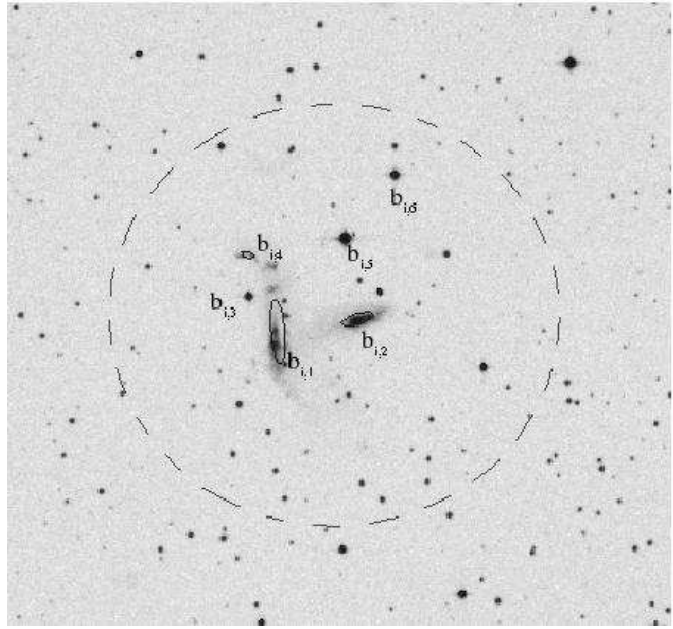}
\caption{An example of a matching problem. An HI detection from the
  HIPASS catalogue is being matched to an optical object in
  SuperCOSMOS.  The object from the sparse HIPASS catalogue $a_i$ is located at the centre 
of this image, the circle represents the
$2 \sigma$ limit of the positional uncertainty.
There are a number of candidate optical counterparts from the denser SuperCOSMOS 
catalogue, $b_{i,1} \dots b_{i,7}$, (circled).  
}
\label{mismatch1}
\end{figure}

In simple terms, our problem is resolving which object in a dense catalogue
is the counterpart for a given object in a sparse catalogue.
Consider a sparse catalogue, $A$, and a dense catalogue, $B$.  
In $A$, we have a sparse object, $a_i$, and we would like to know 
which candidate is the real counterpart for this object in $B$.  
We have the candidates $b_{i,1}, b_{i,2},
b_{i,3}, \dots, b_{i,N_i}$ (Fig~\ref{mismatch1}) (each sparse object $a_i$
may have different numbers of candidates denoted by $N_i$). 
The catalogue parameters associated with source $a_i$ are represented
as a vector $\alpha_i$, and similarly for the parameters of the source
associated with $b_{i,j}$ we use $\beta_{i,j}$.

The cross product $A \times B$ of all possible pairings and consists of the union of
two disjoint sets.  
Formally, $M \subset A \times B$ is the set of all linking
pairs of objects and the remaining pairs of objects are 
non-linking: $U = (A \times B) \setminus M$. 
We now introduce an indicator variable $z_{i,j}$
which is equal to one if, and only if, $(a_i, b_{i,j}) \in M$ and is
zero if $(a_i, b_{i,j}) \in U$. 

\noindent
Our overall aim is to estimate the probability of a match conditioned
on all available information.

\begin{equation}
P(z_{i,j} = 1|\alpha_i,\beta_{i,1}, \beta_{i,2}, \beta_{i,3}, \dots, \beta_{i,N_i}).
\label{genera}
\end{equation}
\normalsize

\cite{sutherland} formulate the overall probability as a normalisation of the
likelihood ratio that each individual candidate is a match.
The likelihood ratio is given as:

\begin{equation}
L_{i,j} = \frac{P(\alpha_i,\beta_{i,j}| z_{i,j} = 1)}{ 
P(\alpha_i,\beta_{i,j} | z_{i,j} = 0)}.
\label{lr}
\end{equation}

\noindent
The overall probability can then be calculated :

\begin{equation}
P(z_{i,j} = 1|\alpha_i,\beta_{i,1}, \beta_{i,2}, \beta_{i,3}, \dots, \beta_{i,N_i}) = \frac{L_{i,j}}{\sum_{i=1 \dots N_i}L_{i,j} + \kappa}.
\label{sns}
\end{equation}
\normalsize

\noindent
We include the \cite{sutherland} justification for this formula in
Appendix~A.  A 
priori we expect that each sparse object is likely to match to some
dense candidate.  If we know the number of candidates then our belief
in this is altered.  The probability that the dense object has a match
when there are $N$ candidates (which we do not know the parameters of)
is given by $\frac{1}{N + \kappa}$.  The $\kappa$ parameter allows for
a probability to be assigned to the state that there is no match for
$\kappa > 0$.  

An independence assumption will hold under a wide range of
circumstances (that is, the probability of the optical properties of a
background object are independent of the properties of the nearest 
radio source).  This allows us to write Equation 2 as :

\begin{equation}
L_{i,j} = \frac{P(\alpha_i,\beta_{i,j}| z_{i,j} = 1)}{ 
P(\alpha_i|z_{i,j}=0)P(\beta_{i,j} | z_{i,j} = 0)}.
\label{lr}
\end{equation}

\noindent
A further assumption that position is independent of other parameters
may also be reasonable, giving :

\begin{equation}
L_{i,j} = \frac{f(\Delta RA, \Delta Dec)}{k}
\frac{P(\alpha'_i,\beta'_{i,j} | z_{i,j} = 1)}{P(\alpha'_i|z_{i,j}=0)P(\beta'_{i,j} | z_{i,j} = 0)}.
\label{lr2}
\end{equation}
\normalsize

Here $f(\Delta RA,\Delta Dec)$ is the probability density function
(pdf) on positional uncertainty 
and $k$ is the density of background objects per unit area, where 
$\alpha'_i$ refers to the non-positional parameters of $\alpha_i$ and,
$\beta'_{i,j}$ refers to the non-positional parameters of $\beta_{i,j}$.

Using the \cite{sutherland} approach the problem is solved if we obtain estimators
for $P(\alpha_i,\beta_{i,j}|z_{i,j}=1)$ and
$P(\alpha_i|z_{i,j}=0)P(\beta_{i,j}|z_{i,j}=0)$.  However in
\citet{rohde2,rohde} a decision rule based on a thresholding
of $P(z_{i,j}=1|\alpha_i, \beta_{i,j})$ is instead used.  In the
arguments that follow we show how to combine estimators of this
discriminative form
to produce an estimator conditioned on all available information as in
Equation~\ref{sns}.  We discuss possible advantages of this approach later.

In this paper we find it useful to convert between probability and
a likelihood ratio.  The likelihood ratio is the probability of the
data given that the object is a match divided by the probability of
the data given it is a non-match.  In this case the likelihood ratio
is equivalent to the posterior odds of class membership. Odds is
related to probability by $o = \frac{p}{1 - p} = \frac{1}{p^{-1} -
  1}$, where $p$ is 
probability and $o$ is the odds.  The likelihood ratio $L_{i,j}$ is
equivalent to the odds of $b_{i,j}$ being a link in the absence of
candidate information, divided by the prior
odds ($\frac{P(z_{i,j}=1)}{P(z_{i,j}=0)} $).  The \cite{sutherland} result presented
here in
Equation~2 and 3 is an expression of the \emph{probability} that $a_i$
links to $b_{i,j}$ given all information.  The probability is calculated as the normalisation of the
\emph{odds} of each 
candidate in ignorance about candidate information.  The conversion to
odds given a probability is given by 
$L_{i,j} = \frac{P(\alpha_i,\beta_{i,j} | z_{i,j} = 1)}{P(\alpha_i,\beta_{i,j} | z_{i,j}=0)} = 
\frac{ \frac{P(z_{i,j}=0)}{P(z_{i,j}=1)}}{P(z_{i,j}=1|\alpha_i,  \beta_{i,j})^{-1} - 1}$
(This can be verified by
substituting Bayes' rule for $P(z_{i,j}=1|\alpha_i,\beta_{i,j})$).  Applying the \cite{sutherland} result we find:

\begin{equation}
P(z_{i,j}=1 | \alpha_i,\beta_{i,1},\beta_{i,2}, \cdots, \beta_{i,N}) = \frac{
\frac{\frac{P(z_{i,j}=0)}{P(z_{i,j}=1)}}{P(z_{i,j}=1 | \alpha_i,\beta_{i,j})^{-1}-1}
}{
\sum_{k=1..N}\frac{\frac{P(z_{i,j}=0)}{P(z_{i,j}=1)}}{P(z_{i,k}=1|\alpha_i,\beta_{i,k})^{-1}-1} + \kappa
\label{full2}
}
\end{equation}

\noindent
which simplifies to

\begin{equation}
P(z_{i,j}=1 | \alpha_i,\beta_{i,1},\beta_{i,2}, \cdots, \beta_{i,N}) = \frac{
\frac{1}{P(z_{i,j}=1 | \alpha_i,\beta_{i,j})^{-1}-1}
}{
\sum_{k=1..N}\frac{1}{P(z_{i,k}=1 | \alpha_i,\beta_{i,k})^{-1}-1} +
\frac{P(z_{i,j}=1)}{P(z_{i,j}=0)} \kappa.
\label{full}
}
\end{equation}

\noindent

Using a discriminative probabilistic classifier, it is
possible to estimate $P(z_{i,j}=1 | \alpha_i,\beta_{i,j})$, and using
Equation ~\ref{full}, to combine all candidate information to arrive at a
probability of a match conditioned on all available information.  We
use this rule to calculate probabilities using the discriminative
method below.

While we would normally estimate $P(z_{i,j}=1 | \alpha_i,\beta_{i,j})$
directly it is useful to consider the component probabilities to this estimator.

\begin{equation}
P(z_{i,j}=1|\alpha_i, \beta_{i,j}) = \frac{P(\alpha_i,
  \beta_{i,j}|z_{i,j}=1) P(z_{i,j}=1)  }{P(\alpha_i,
  \beta_{i,j}|z_{i,j}=1) P(z_{i,j}=1) + P(\alpha_i,
  \beta_{i,j}|z_{i,j}=0)P(z_{i,j}=0)}.
\label{mlestimate}
\end{equation}

\noindent
It is a well justified assumption for our problem that $P(\alpha_i,
\beta_{i,j}|z_{i,j}=0) = P(\alpha_i|z_{i,j}=0)
P(\beta_{i,j}|z_{i,j}=0)$.  However this is not taken into account
when $P(z_{i,j}=1 | \alpha_i,\beta_{i,j})$ is estimated using a
standard algorithm. This assumption is not utilised \citet{rohde2,rohde}.

\subsection{The Two Approaches to Classification}

Classification problems can be approached in two different ways.  The
first is to estimate the probability $P(C=1|x)$\footnote{$P(C=1|x)$ is
the probability that $x$ belongs to class $1$ conditioned on $x$.}.  We describe this as
the discriminative approach.  Sometimes the word discriminative is
used to refer to a decision boundary however by simple
decision theoretic arguments, this is simply a threshold of
$P(C=1|x)$ \citep{Duda_Hart_Stork}.  Here we use the word discriminative more generally to
refer to $P(C=1|x)$.  Discriminative classification is the basis of
many methods including neural networks,
Platt calibrated \citep{platt} Support Vector Machines (SVMs)
\citep{vapnik} and logistic regression \citep{hosmer}.

The alternative is to calculate the component probabilities $P(x|C=1)$
and  $P(x|C=2)$ and $P(C=1)=1-P(C=2)$ and then apply
Bayes' rule in order to obtain $P(C=1|x)$.  We will describe such techniques as
generative methods.  This approach is used in,
for example, nearest neighbours' approaches to classification.

In machine learning it is more common to estimate $P(C=1|x)$ rather
than all the component probabilities.  \citet{vapnik} argues that
using the (simple) discriminative formulation is a fundamental principle
of statistics.  \citet{hand} remarks that discriminative models
such as neural networks 
  are successful because they find an intermediate position between
  simple parametric models and complex generative density estimation
  models such as nearest neighbours \footnote{Nearest neighbours is a
    conceptually simple algorithm however it has a large number of
    effective parameters (proportional to the number of datapoints)
    this can lead to complex decision boundaries.}.

The above framework for matching allows the probability of a match
to be calculated from either generative or discriminative estimators.
Competing arguments for the merits of each approach exist:

\begin{enumerate}
\item
The generative model $P(x|C=1)$, $P(x|C=2)$ and
$P(C=1)=1-P(C=2)$ allows the likelihood ratio method to be used
directly Equations (2-5).  The simplifying independence assumption can
be introduced into Equation 4.  
\item
The discriminative model $P(C=1|x)$ is in agreement with the
\emph{principle} suggested in \citet{vapnik} and takes the intermediate
complexity which is argued as positive in \citet{hand} (between
parametric and nearest neighbour methods), and is generally the 
more common machine learning approach.  However the independence assumption in Equation 8 must be ignored and the model is left underconstrained.
\end{enumerate}

In this paper we apply both a generative and a discriminative model.  The generative density estimation is
performed by 
using a high dimensional Gaussian Mixture Model fitted to the data
using the Expectation Maximisation (EM) algorithm.  This produces estimates of $P(\alpha_i,
\beta_{i,j} | z_{i,j}=1)$ and
$P(\alpha_i|z_{i,j}=0)P(\beta_{i,j}|z_{i,j}=0)$.  The \cite{sutherland} formalism can
be applied by using Equation ~\ref{sns}  and Equation ~\ref{lr}.

The discriminative model is fitted using an SVM with Platt Calibration
\citet{platt}.  This produces an estimate of
$P(z_{i,j}=1|\alpha_i, \beta_{i,j})$ which can be combined with the
probabilities of all the other candidates ($b_{i,1}, \cdots,
b_{i,N_i}$) using Equation ~\ref{full}.

\section{Algorithms}
\label{algorprinc}
A number of different non-parametric approaches are available for 
the problems of density estimation and estimating class probabilities.
Both approaches are dominated by the use
of the principle of maximum likelihood. 

\subsection{Density Estimation}

According to the formulation in Section 2, our problem is solved if we obtain a good estimator of the densities
$P(\alpha_i, \beta_{i,j}|z_{i,j}=1)$ and $P(\alpha_i|z_{i,j}=0)
P(\beta_{i,j}|z_{i,j}=0)$.

One method for density estimation is the use of
the k-nearest neighbours averaging.  This method has a very large
number of effective parameters that increase in proportion to the size
of the dataset.  The k-nearest neighbours averaging is very
discontinuous which seems undesirable.  The kernel density estimation
technique overcomes this by smoothing the output using a Gaussian
kernel.  The kernel method also has a very large number of effective
parameters which means there is a high chance of producing an overly
complex model that fits the idiosyncrasies of the data, this is of
particular concern when applying the method to high dimensional
problems.  Discussion and comparison of these techniques can be found
in \citet{hastie}.

The most appropriate method for this is to fit a semi-parametric
model such as a Gaussian mixture model to the data using maximum
likelihood.  This can be achieved using the Expectation Maximisation (EM) algorithm.  The EM
algorithm allows the analytical form for maximum likelihood estimators
for mean and standard deviation to be used on a mixture model by
introducing the concept that every mixture component has a certain
responsibility for every data point.  The E step calculates the
responsibilities where the M step maximises the likelihood \citep{mclachlanem}.  EM is an
iterative hill climbing algorithm on the likelihood function.
In this paper we use the EM implementation provided by the
Netlab software package \citep{Nabney}.  It is possible for the EM
algorithm to converge to a poor local optimum of the likelihood
function.  For this reason
it is often necessary to attempt the optimisation with different
initial conditions.  We follow the recommendation in \citet{Nabney} to use the
k-means clustering algorithm in order to set the initial parameters of the EM algorithm.

\subsection{Class Probabilities}

Neural networks are multi-parameter models that can be fit to labelled
data using the
principle of maximum likelihood to estimate class probabilities.  When
a neural network is fit to targets that are binomial class
probabilities, the use of a `cross-entropy' error function leads to a
maximum likelihood estimate of the class probability \citep{bishop}.
In previous work we found that neural networks showed sub-optimal performance for
achieving high classification rates on the SuperCOSMOS-HIPASS matching
problem \citep{rohde2}.

Support Vector Machines (SVMs) are primarily classification
algorithms.  The SVM projects the input vectors into a high
dimensional feature space and finds a separating hyperplane.  The
separating hyperplane is chosen using the criterion of maximal margin
which means that the distance of the plane to the closest points (of
opposing classes) is maximised \citep{cristianini} \citep{vapnik}.  In
this study we use the SVM lite software \citep{joachims} in order to
fit the model.
The SVM output is a real number ($g$), with the 
sign representing the side of the plane and the magnitude represents the
distance from the plane.  A binary classifier is produced by placing a
threshold of zero on $g$.  A probabilistic classifier can be obtained
by considering $P(C=1|g)$.  We follow the method proposed in
\citet{platt} where a logistic sigmoid is fitted to $P(C=1|g)$ using
the principle of maximum likelihood in order to obtain probabilistic
outputs -

\begin{equation}
P(C=1|g) = \frac{1}{1 + e^{w_1 g + w_2}}.
\label{sigmoid}
\end{equation}

The parameters $w_1$ and $w_2$ are adjusted as part of fitting the
model, in order to maximise the
likelihood of the sigmoid.  We again use the Netlab package
to do this using a quasi-Newton optimisation algorithm.

The details of this calibration process are as follows.  
Platt calibration involves a three-fold-cross-validation procedure for
fitting the sigmoid.  The SVM is fitted to two thirds of the training
data, the remaining third is used to determine $w_1$ and $w_2$.  This
procedure is repeated three times and the average value of $w_1$ and
$w_2$ is then used.  In order to avoid a very rapid transition of the
output probability from zero to one, Platt recommends training on
non-binary targets.  Hence, rather than have training points labelled
as $1$
for $C=1$ and $0$ for $C=2$, they have non-binary values.  Instead
we use $\frac{M_1+1}{M_1+2}$ for $C=1$ ($M_1$ is the number of training
examples drawn from class $1$) and $\frac{1}{M_2+2}$ for $C=2$ ($M_2$
is the number of examples drawn from class $2$).  This is 
justified using regularisation arguments in \citet{platt}.

\section{Application of methods to SuperCOSMOS \& HIPASS}
\label{applic}
The matching of the HIPASS radio catalogue \citep{meyer} to the
optical SuperCOSMOS catalogue \citep{hambly} is a difficult problem due
to the poor positional uncertainty on the HIPASS
catalogue ($\sigma \approx 1$ arcmin).  The use of external redshifts from the NASA Extragalactic
Database and 
the Six Degree Field Survey \citep{Wakamatsu} along with human judgment are
able to match approximately 
half of this catalogue known as HOPCAT (HIPASS Optical Catalogue) \citep{doyle}.  In previous work a
(non-probabilistic) binary classifier was applied to this training
data \citep{rohde}.  A cross-validation process gave very good overall
performance $99.12$ per cent.  However this involved only applying the
classifier to a single case at a time, we however made the additional
assumption that exactly one candidate could be a match.  The binary
SVM however did not incorporate this assumption.
On unmatched date the binary SVM found $1209$ new matches.  In $1012$
other cases the classifier gave an ambiguous result either selecting zero or multiple
matches.  In this section we demonstrate the ability of probabilistic approaches
to enforce a constraint that every sparse object
has exactly (or at most) one match.

\subsection{Model Validation}
\label{validation}
We have available two broad approaches for calculating probabilities.
How can we then discriminate between which is the better of the two
approaches?  What makes a probability a good probability?  This turns
out to be a very difficult philosophical question.  The definition of
probability is the source of one of the most celebrated disputes in
science \citep{howie} and results in two competing paradigms for
statistics, frequentist and Bayesian.

Rather than delving deeper into this philosophical issue we offer
some heuristic tests to evaluate the quality of probability.
The fact that a number of intuitive tests exist results from the fact
that there is no clear cut method for evaluating the quality of the
probabilities produced.  We present the measures for the three methods
considered, but the nature of the problem only allows us to make
intuitive statements about which method is better.

It is standard practice to put part of the data aside in a test set in
order to evaluate properties of the model.  Our model validation
involves training on $75$ per cent of the data, $9941$ training vectors (i.e. 
$1356$ positive examples and $8585$ negative examples) and testing on
the remaining $25$ per cent $3304$ vectors ($453$ positive and $2851$ negative examples).

The measures of probability quality that we consider are:
\begin{enumerate}
\item
Classification rate or percentage correctly classified.  One of the main reasons for producing
a probability is to correctly match our datasets, testing the
classifiers ability to correctly classify a test set has obvious
intuitive appeal.  This test however gives equivalent results for
any monotonic increasing function of the probability.  We make this
test by counting how often the classifier assigns the highest
probability to the correct match on our test set.
\item
Calibration is the property that if all of the examples where
$P(C=1|x)\approx k$ are binned then the proportion of objects in that bin
belonging to class one should be approximately $k$.  For example if
$P(C=1|x) = 0.2$ then the classifier is 
well calibrated when $20$ per cent of objects with this value of $x$ belong to
class $1$ .  Calibration is only part of the picture as perfect calibration can
arise if the classifier makes dishonest predictions in order to
obtain frequencies in agreement with predicted probabilities
\citep{degroot}.  The property of calibration is also independent of
classification rate. 
\item
A number of scoring rules have been introduced in order to rank
probabilities.  The most widely used is the \citet{brier} score which
is a mean squared error statistic.  For example if the classifier
predicts a value of $0.9$ and it is a match, then the contribution to
the Brier score is $0.1^2=0.01$.  If for example it is not a match, then
the contribution is $0.9^2=0.81$.
\end{enumerate}

Note that minimising (i) and (iii) alone can result in overfit models
that are overly complex and generalise poorly.  Calibration becomes
important when there is an inevitable proportion that is
misclassified.  In general minimising the classification rate will
also cause the Brier score to be lowered.

One of the advantages of using probabilistic outputs is that it allows for
cases where it is difficult to determine a match to be discarded.  
In pattern classification the Receiver Operator Characteristic (ROC) curve
is commonly used in order to 
describe the trade-off between incorrect classifications, false
positives and false negatives \citep{Duda_Hart_Stork}.  False
  positives are negative examples that have been incorrectly labelled
  positive, and similarly false 
  negatives are positives that have been incorrectly labelled
  negative.  This can be directly interpreted as a trade-off
between completeness and efficiency where completeness is the
proportion of sparse objects one finds matches for, and efficiency is
the number of correct matches in the dataset.  The optimal trade-off
will be application dependent.  We see the trade-off for our dataset
for our two models in Fig~2.  A perfect curve would be one where the efficiency was $1$
for every completeness between $0$ and $1$.  From these plots it is
apparent that the SVM achieves better classification than the mixture
model for all possible completenesses.

We evaluate calibration by the use of reliability diagrams \citep{caruana}.
Reliability diagrams are produced by binning the output of the
classifier and testing if the class frequencies (on the test set) are
in agreement with the model's prediction.  The
SVM and mixture reliability diagram is shown in Fig~3. The error bars
are 90 per cent
credible regions.  Credible regions are the Bayesian alternative to
confidence intervals. In our example a clear
advantage is that the credible region takes into account a priori
knowledge that probabilities can only be between $0$ and $1$.
A persuasive argument for credible regions over confidence
intervals is found in \citet{jaynes.intervals}.  Confidence regions are
calculated using a uniform Beta conjugate prior.  See 
\citet{bernardosmith} for details.  The error bar indicates a region
with $90$ per cent probability of containing the frequency with which
probabilities are assigned to a particular class.  The centre point is
the posterior median.  Both the SVM and the mixture
model exhibit satisfactory calibration.  This is demonstrated by the
fact that the probabilities in general fall along the diagonal line.  It is noteworthy that as the
SVM classifies more objects correctly, there is less data available to
calibrate the probabilities.  This causes the error bars to be much
larger for the central probabilities in the SVM reliability diagram
compared to the probability values near zero or one, this is also true
but less pronounced for the mixture model.  

Finally we consider the Brier score which in some way gives a combined
indication about calibration and classification performance.  The SVM
has a Brier Score of $0.0168$; the mixture model has a Brier score of
$0.0863$.  Presumably the SVM's better classification is primarily
responsible for this, at a completeness of $100$ per cent the SVM
correctly classified $94$ per cent where the mixture model classified
$87.5$ per cent correctly.  The benefit of good calibration is only apparent
if errors are unavoidable.

\begin{figure}
  \centerline{\hbox{ \hspace{0.20in} 
\label{roc}
    \epsfxsize=3.0in
    \epsffile{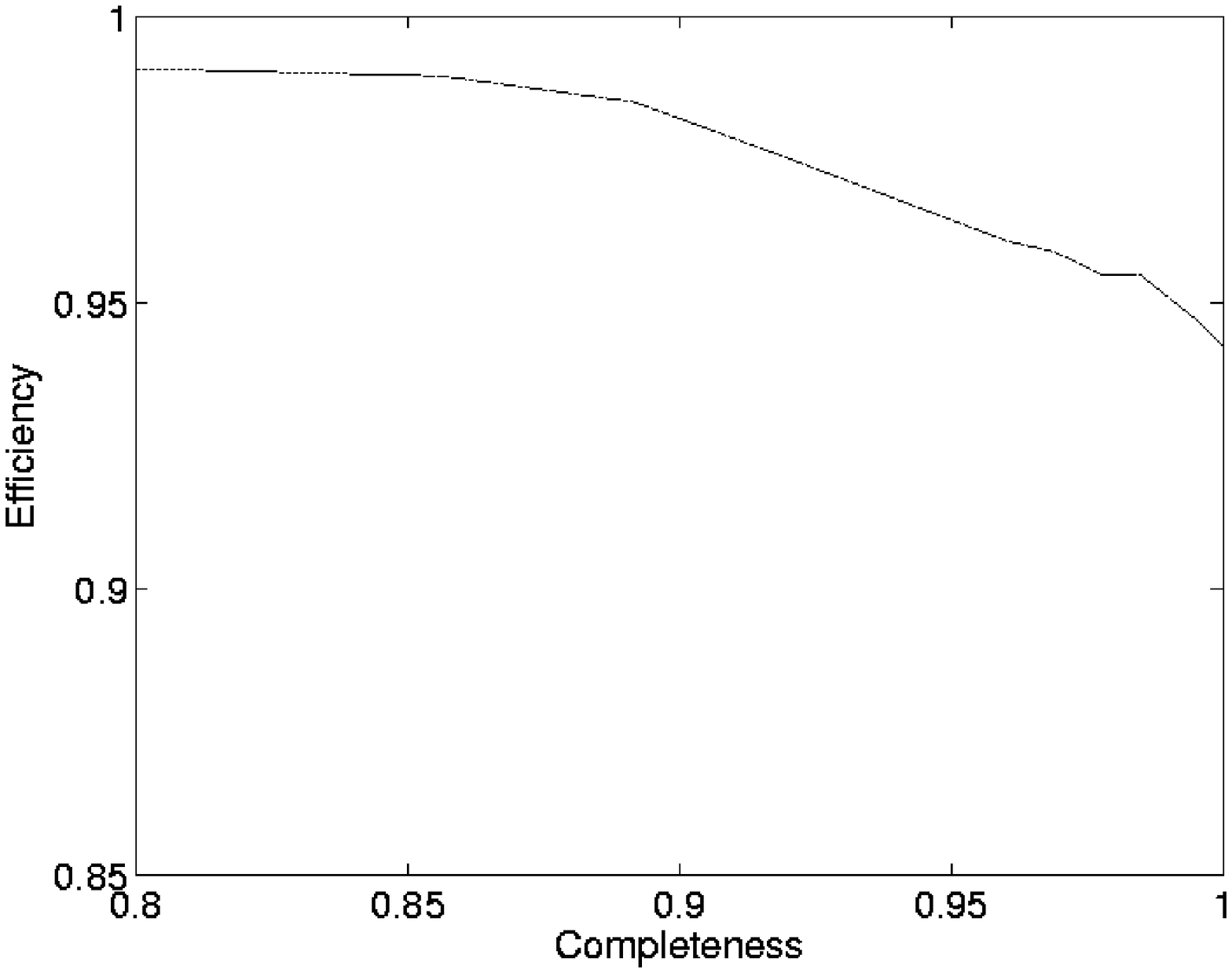}
    \epsfxsize=3.0in
    \epsffile{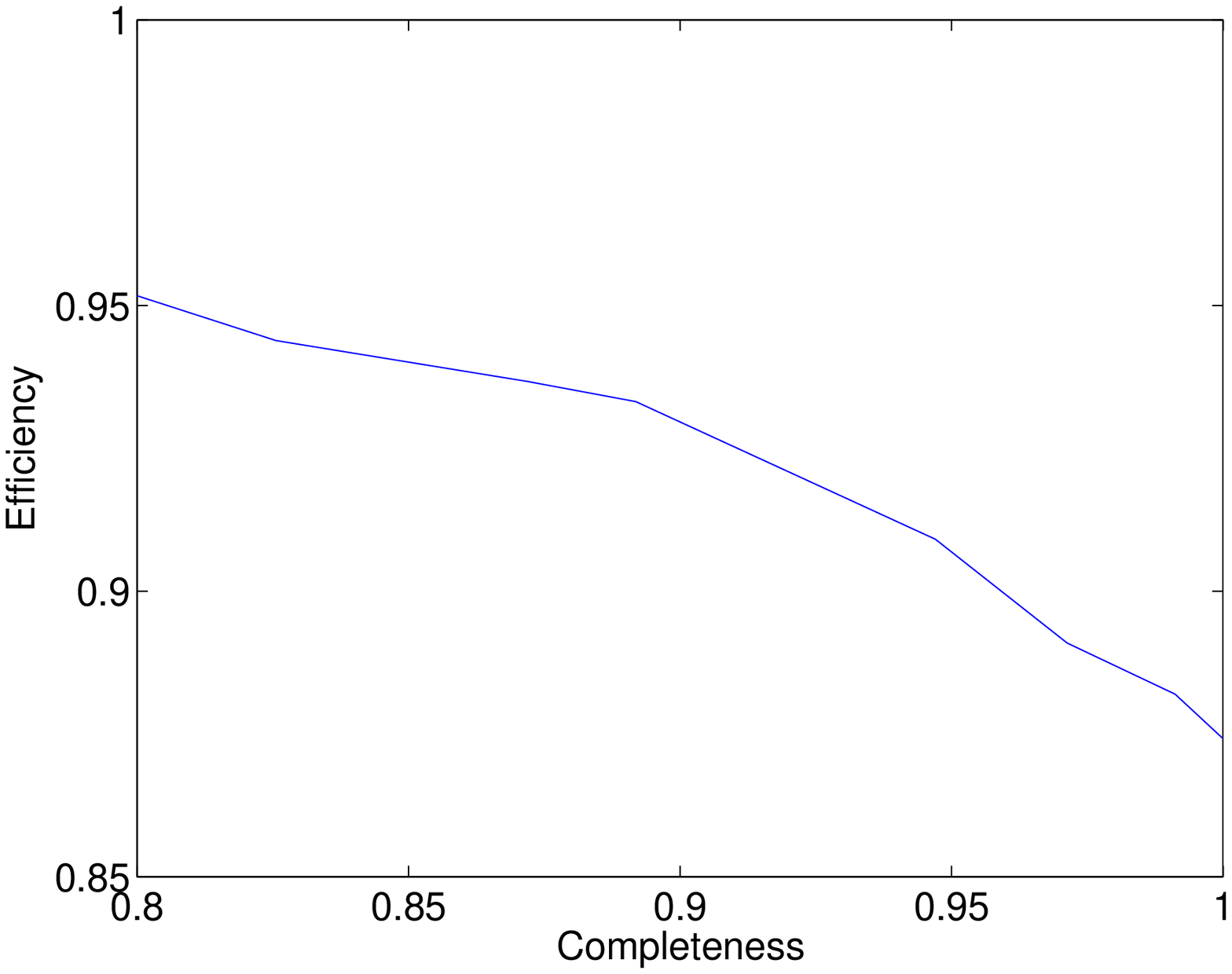}
    }
  }

  \vspace{9pt}
  \hbox{\hspace{0.5in} (a) \hspace{2.8in} (b) } 
  \vspace{9pt}

\caption{Trade-off between completeness and efficiency of (a) the
  discriminative SVM
  based classifier and (b) the generative mixture model based classifier.  The
  SVM based classification is consistently better than the mixture
  based classification.  In practice a decision about the trade-off
  between completeness and efficiency must be made.  While
  completeness can be sacrificed for efficiency dropping below a
  completeness of $0.8$ brings very marginal benefits.  At a
  completeness of $1$ reasonably high efficiencies are obtained.}
\end{figure}

\begin{figure}

  \centerline{\hbox{ \hspace{0.20in} 
    \epsfxsize=3.0in
    \epsffile{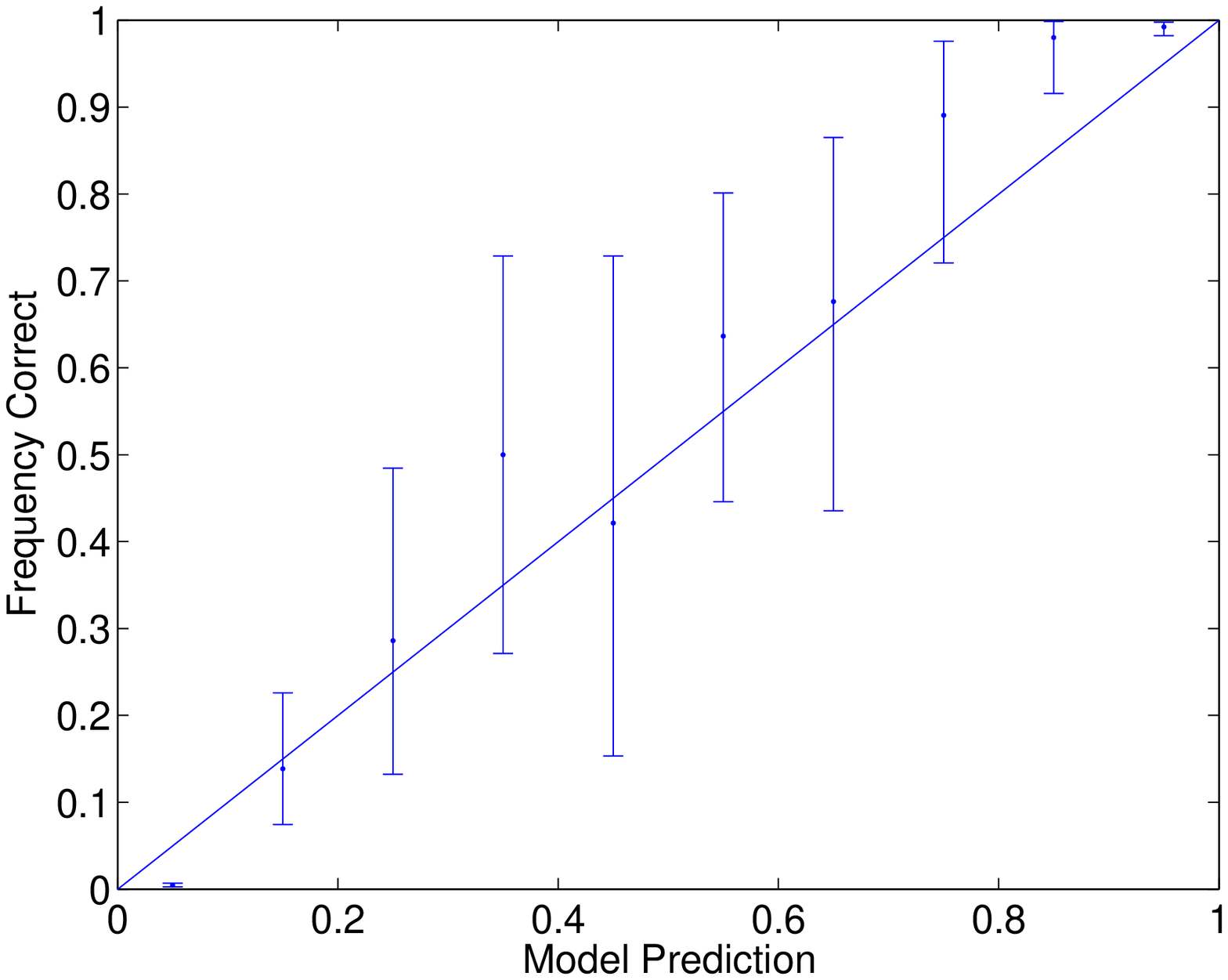}
    \epsfxsize=3.0in
    \epsffile{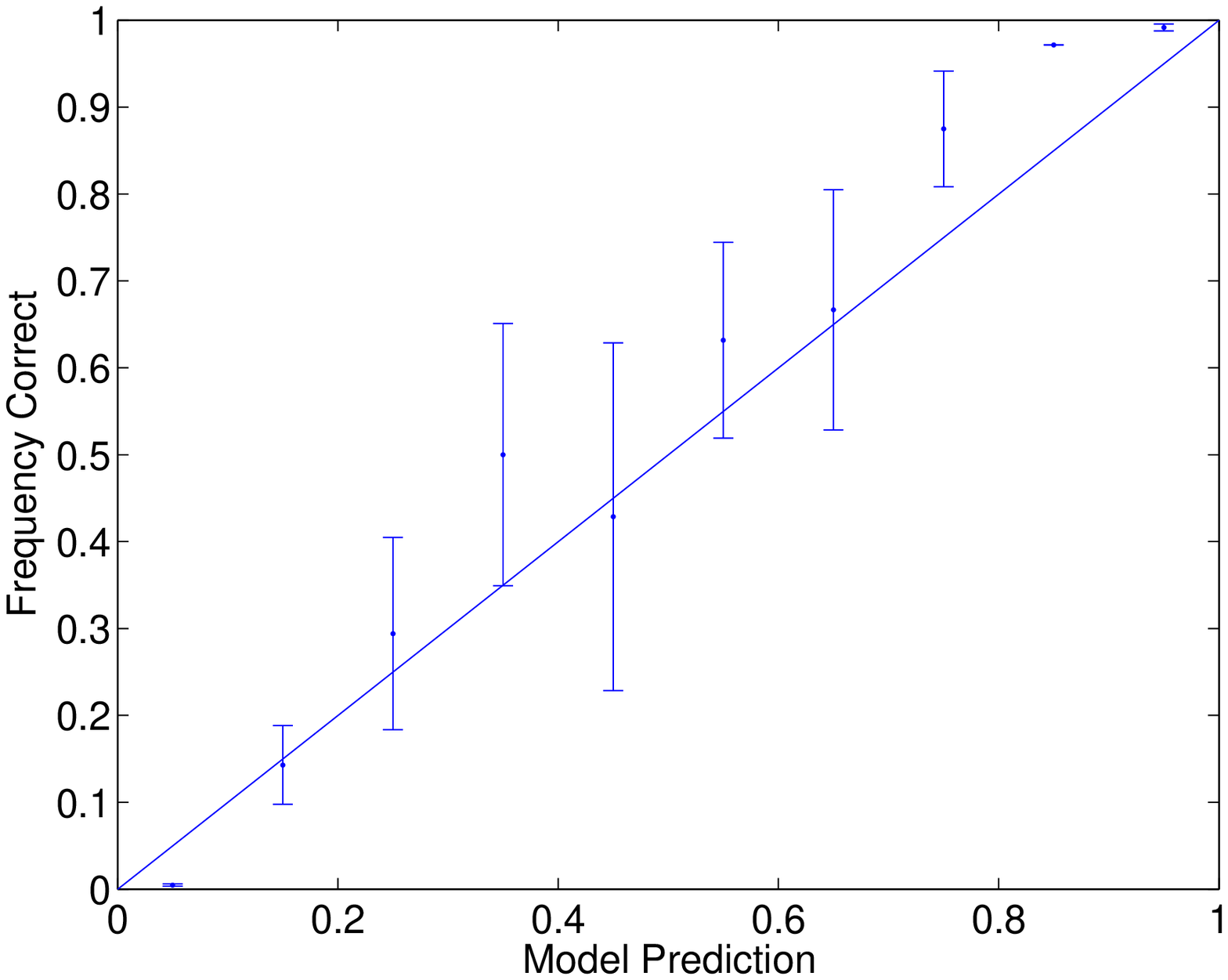}
    }
  }

  \vspace{9pt}
  \hbox{\hspace{0.5in} (a) \hspace{2.8in} (b) } 
  \vspace{9pt}

\caption{Reliability diagrams for (a) the discriminative SVM classifier
  and (b) the generative mixture model classifier.  Good performance is
  indicated by the points lying on or near the diagonal line.}
\end{figure}

\subsection{Examples of Classification}

\begin{figure}
  \centerline{\hbox{ \hspace{0.20in} 
    \epsfxsize=3.0in
    \epsffile{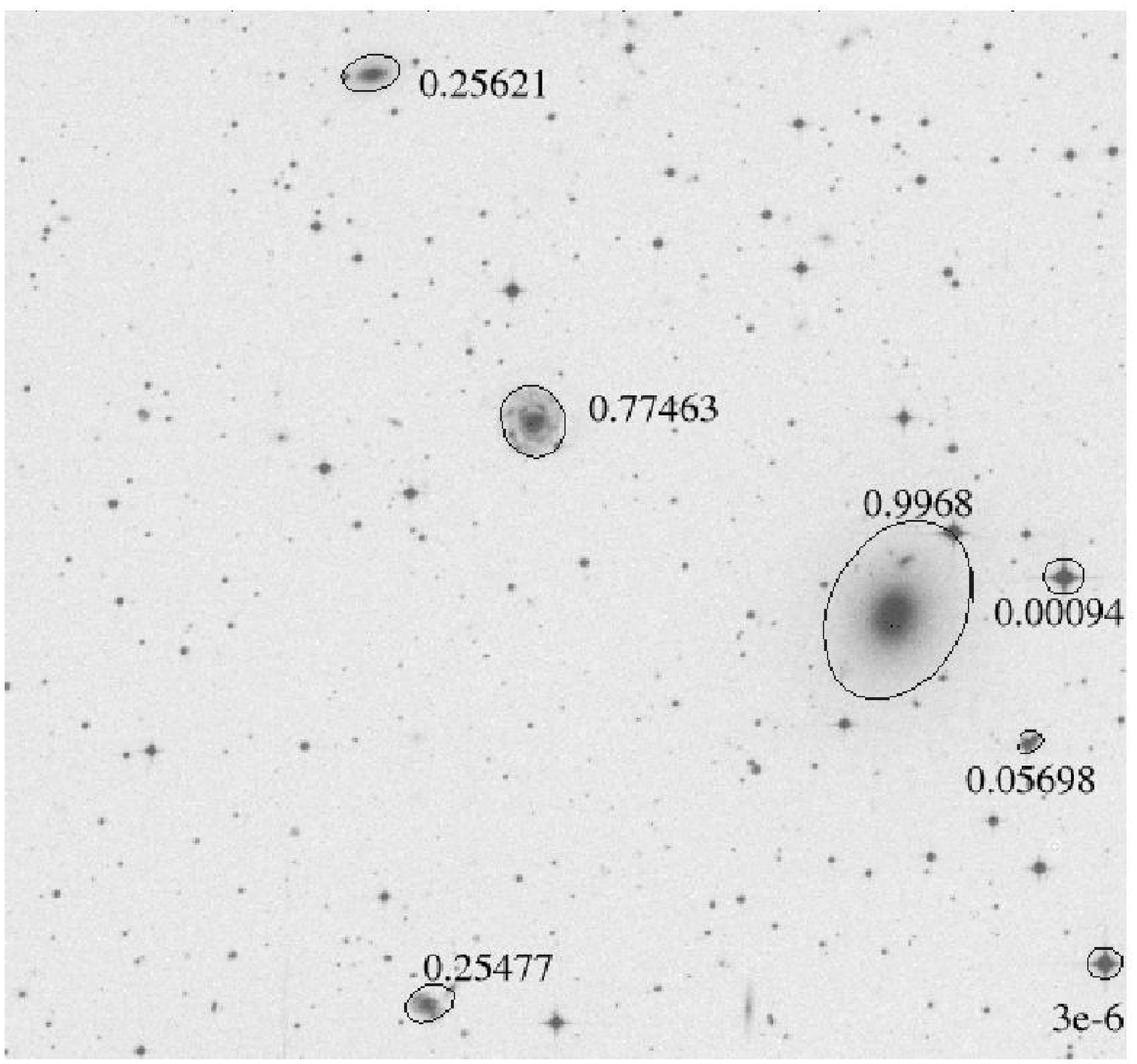}
    \epsfxsize=3.0in
    \epsffile{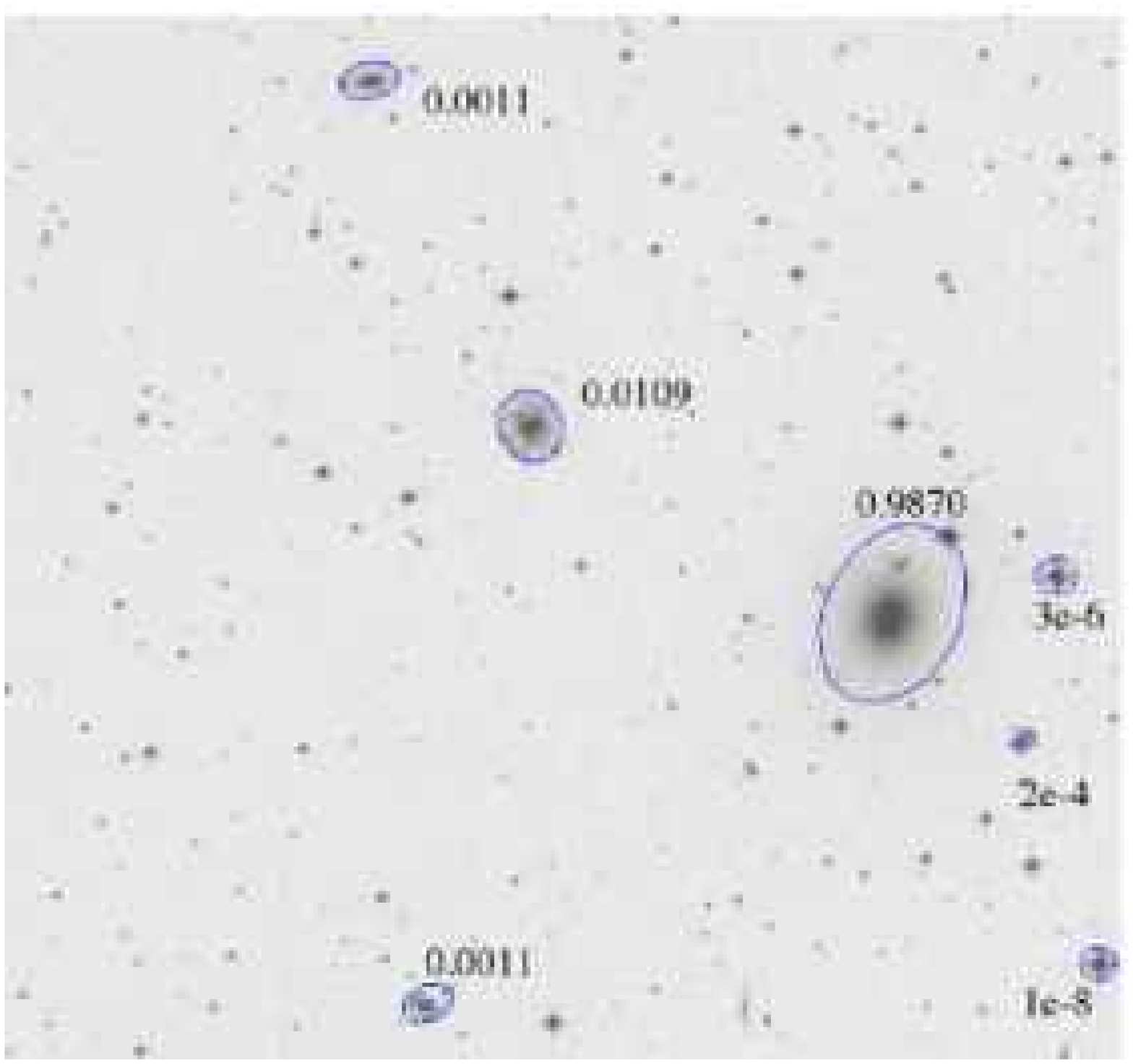}
    }
  }

  \vspace{9pt}
  \hbox{\hspace{0.5in} (a) \hspace{2.8in} (b) } 
  \vspace{9pt}

\caption{The probabilities in the absence of candidate information (a)
  and in the presence of candidate information (b).  Note that candidate
  information makes the probabilities well behaved in the sense that
  they now should sum to one (or less that one if the identification
  rate is less than $1$).}

\label{candinfo}
\end{figure}

\begin{figure}
  \centerline{\hbox{ \hspace{0.20in} 
    \epsfxsize=3.0in
    \epsffile{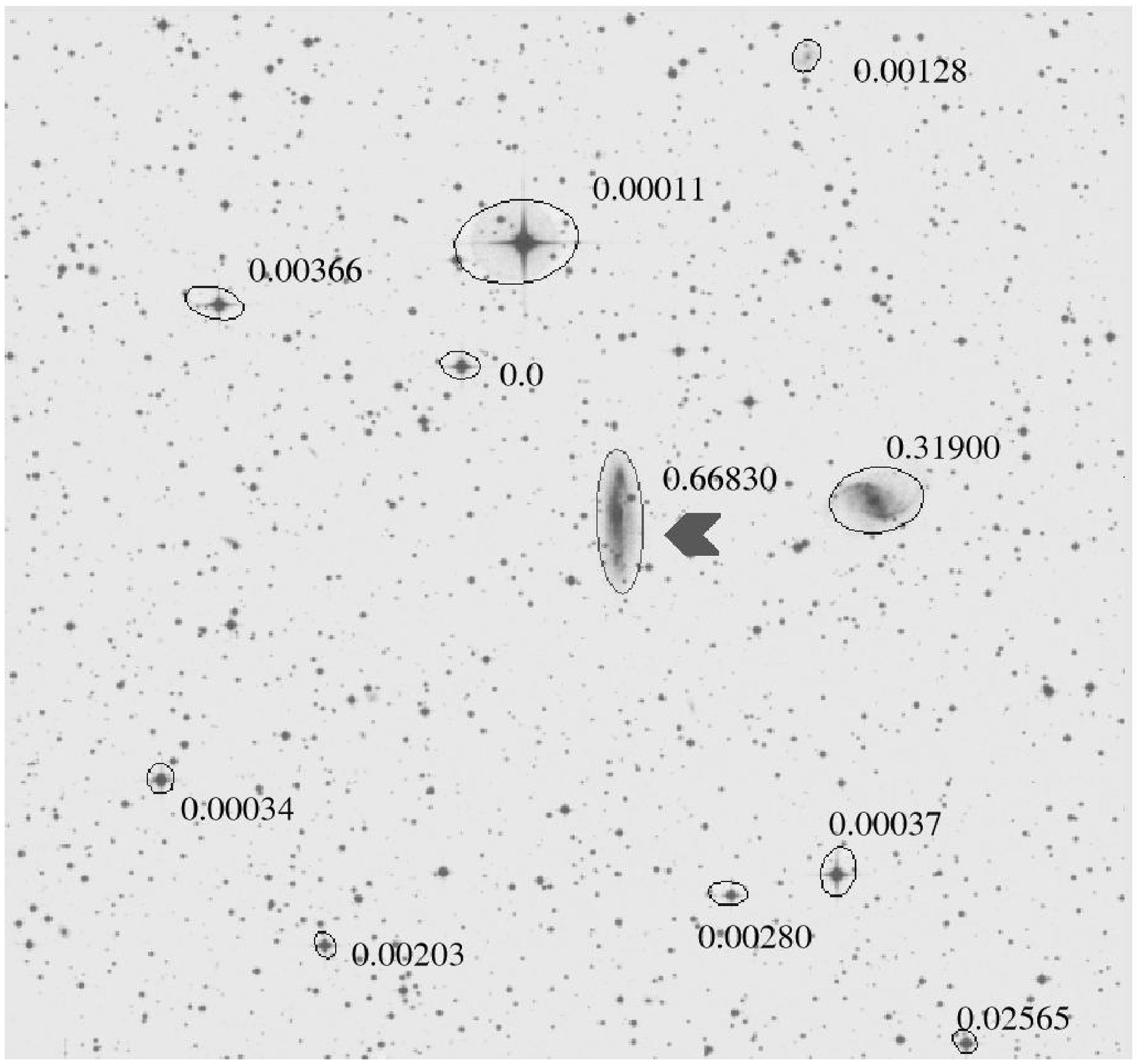}
    \epsfxsize=3.0in
    \epsffile{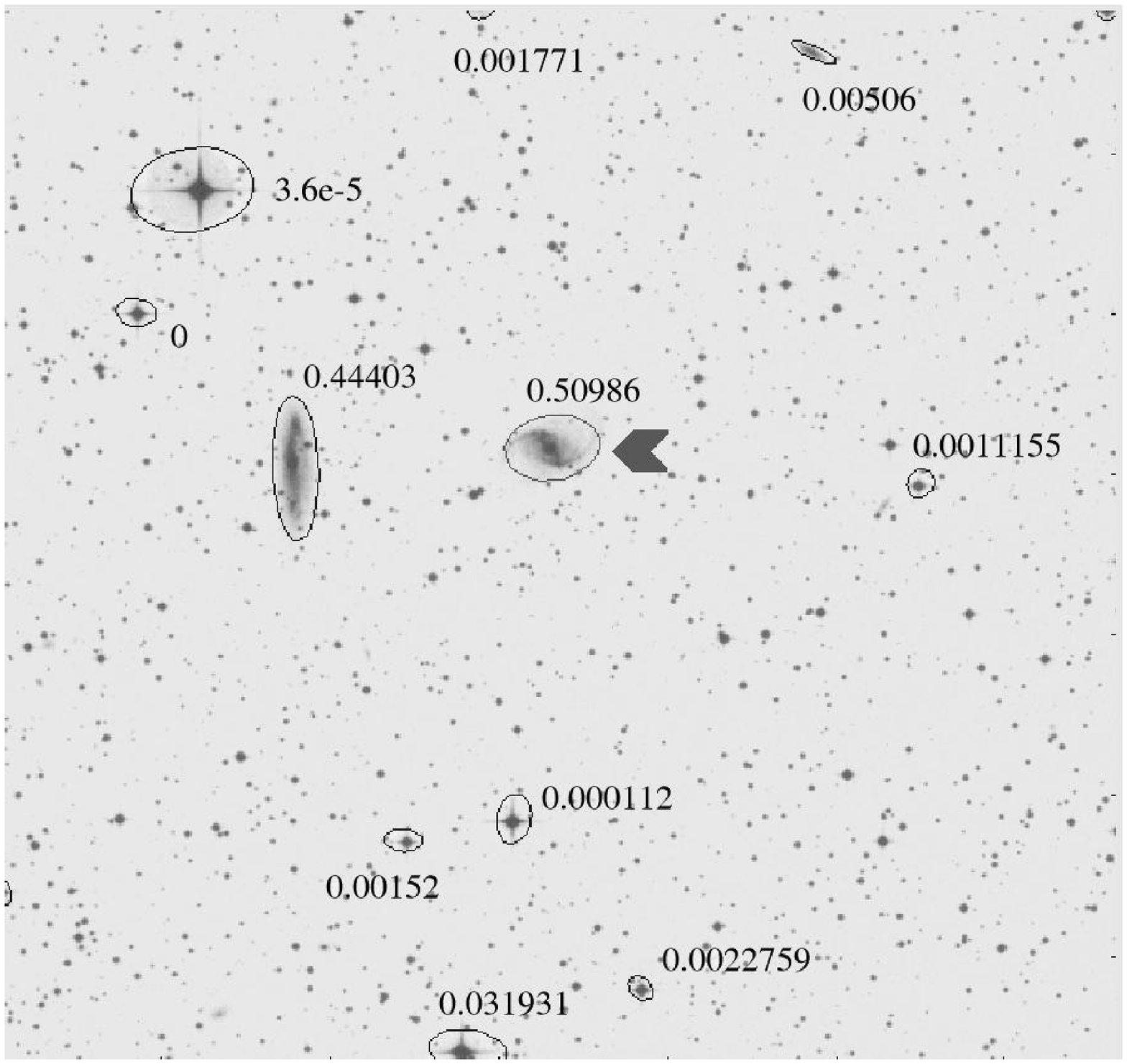}
    }
  }

  \vspace{9pt}
  \hbox{\hspace{0.5in} (a) \hspace{2.8in} (b) } 
  \vspace{9pt}
  \centerline{\hbox{ \hspace{0.20in} 
    \epsfxsize=3.0in
    \epsffile{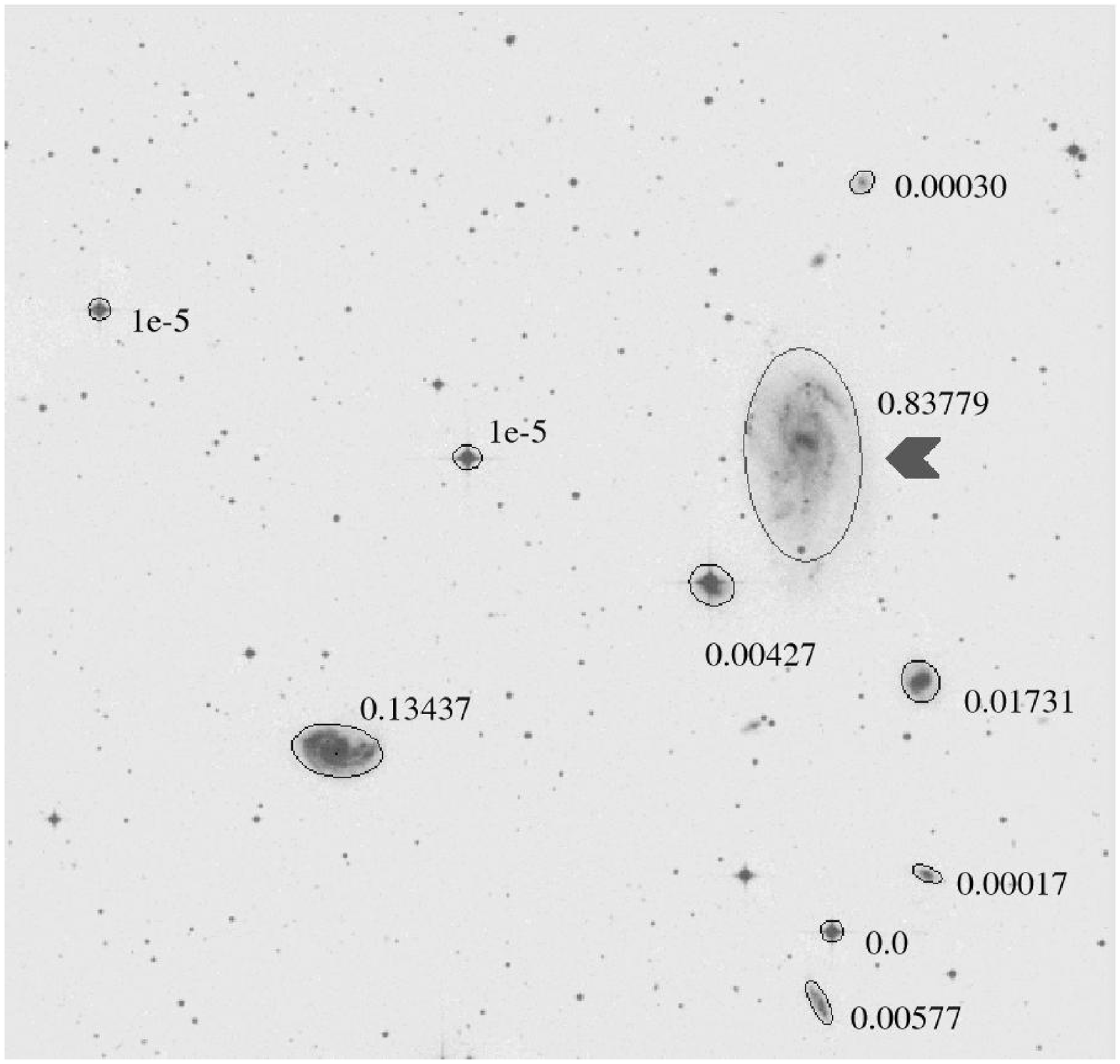}
    \epsfxsize=3.0in
    \epsffile{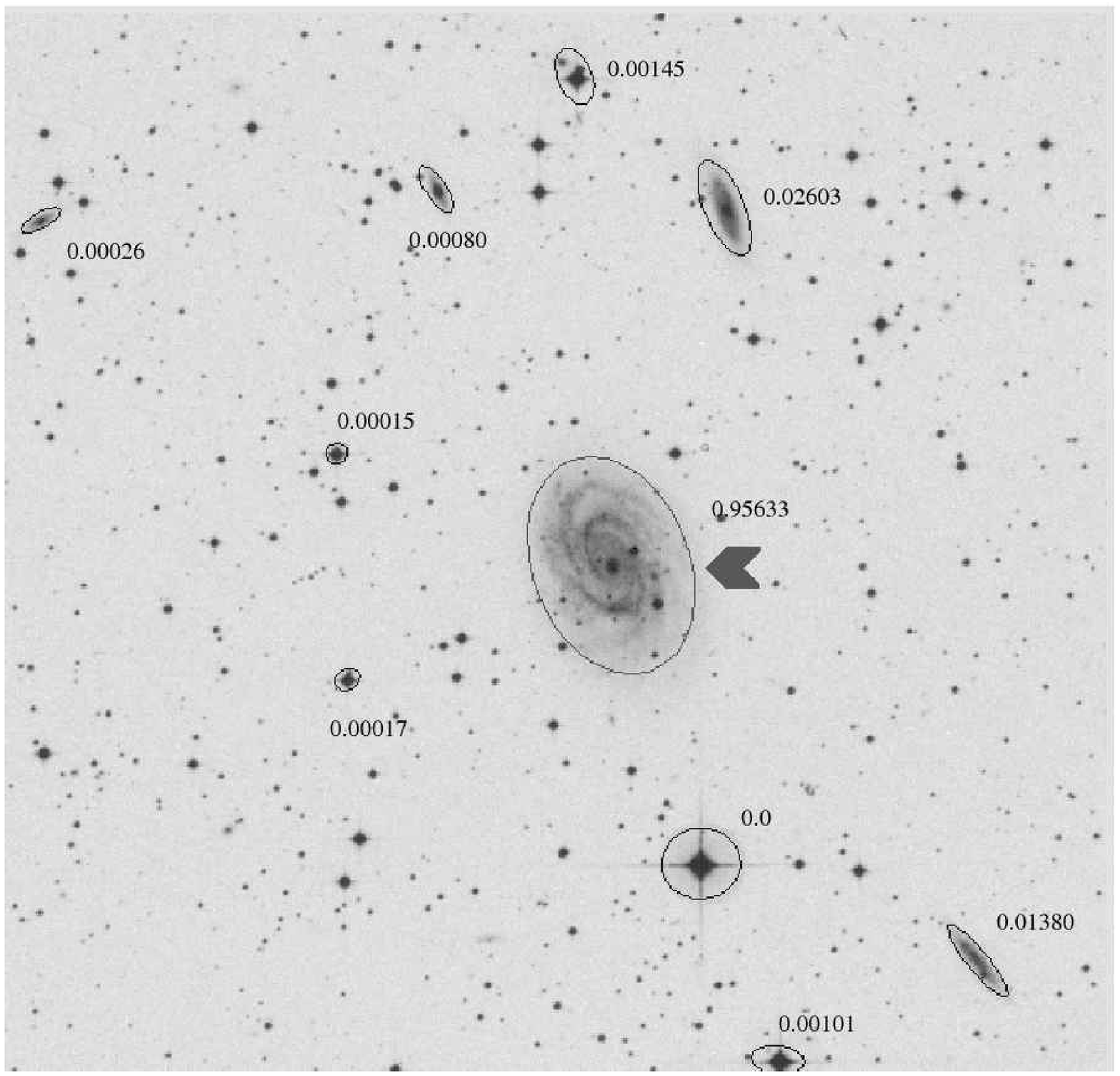}
    }
  }

  \hbox{\hspace{0.5in} (c) \hspace{2.8in} (d) } 
  \vspace{9pt}

  \vspace{9pt}

  \caption{Some examples of correct matches (match marked with an arrow).  
  In each case optical
  redshifts are able to select the matching optical galaxy and
  eliminate the non-linking optical galaxies.  Two HIPASS 
  detections are in close proximity in (a) and (b) and the classifier is
  able to classify correctly, but more importantly offer considerable
  qualification to the classification.  In (c) again we obtain a
  correct match with an understandable level of qualification because of
  the other significant candidate.  In (d) the classifier chooses the
  correct link confidently deciding that the number of background
  galaxies are relatively poor candidates.}

\label{correctexamples}
\end{figure}

\begin{figure}
  \centerline{\hbox{ \hspace{0.20in} 
    \epsfxsize=3.0in
    \epsffile{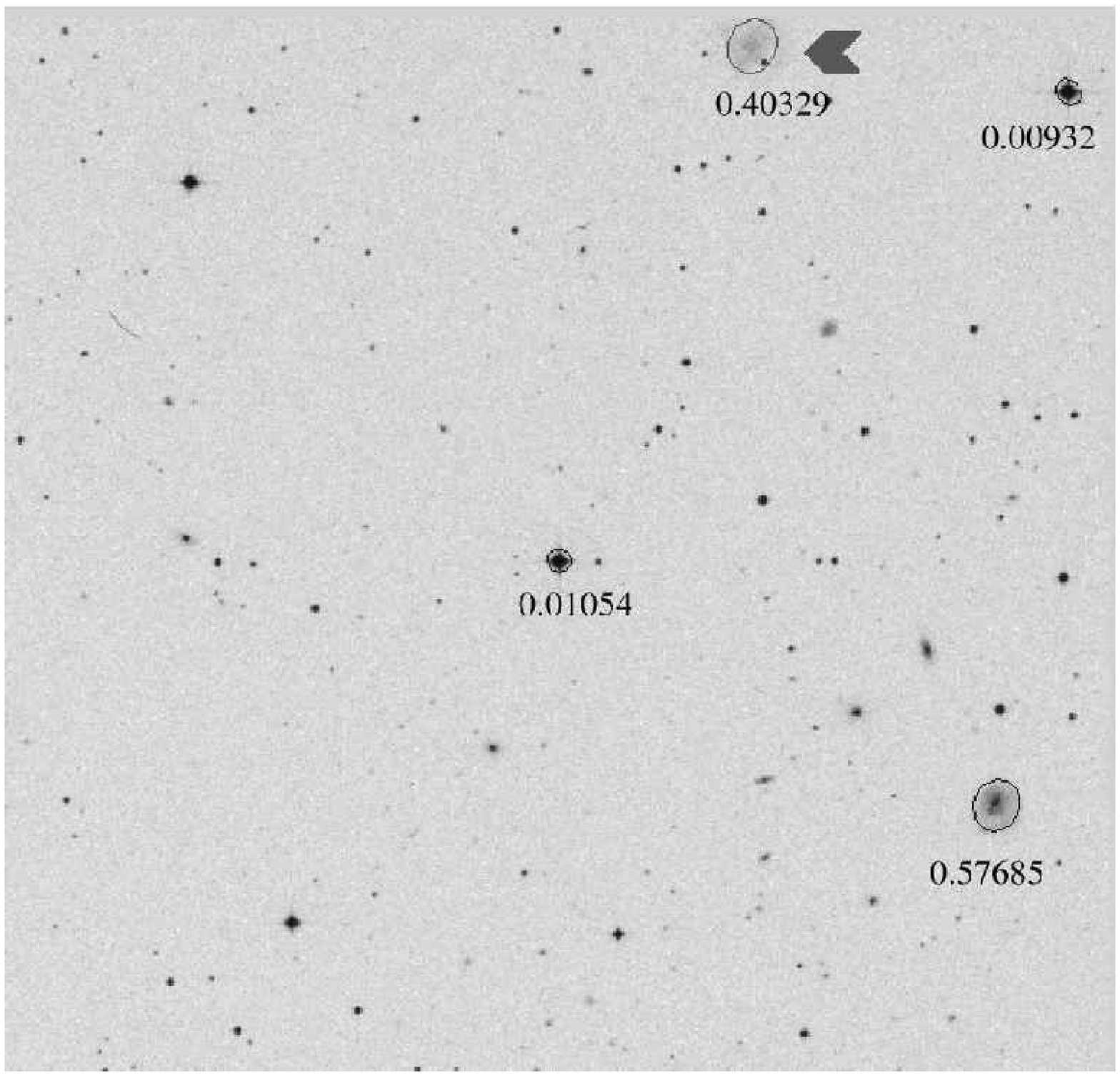}
    \epsfxsize=3.0in
    \epsffile{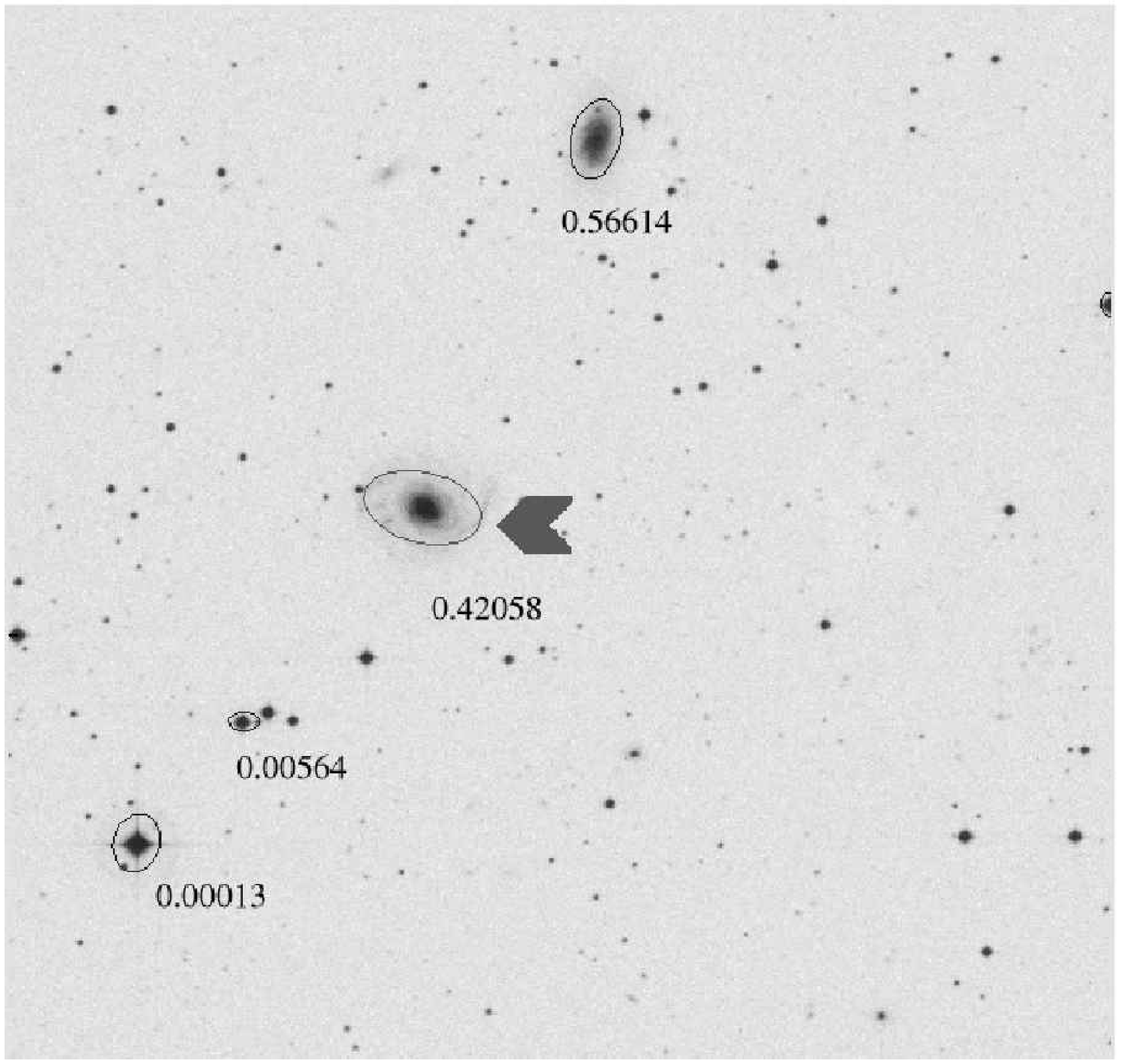}
    }
  }

  \vspace{9pt}
  \hbox{\hspace{0.5in} (a) \hspace{2.8in} (b) } 
  \vspace{9pt}
  \centerline{\hbox{ \hspace{0.20in} 
    \epsfxsize=3.0in
    \epsffile{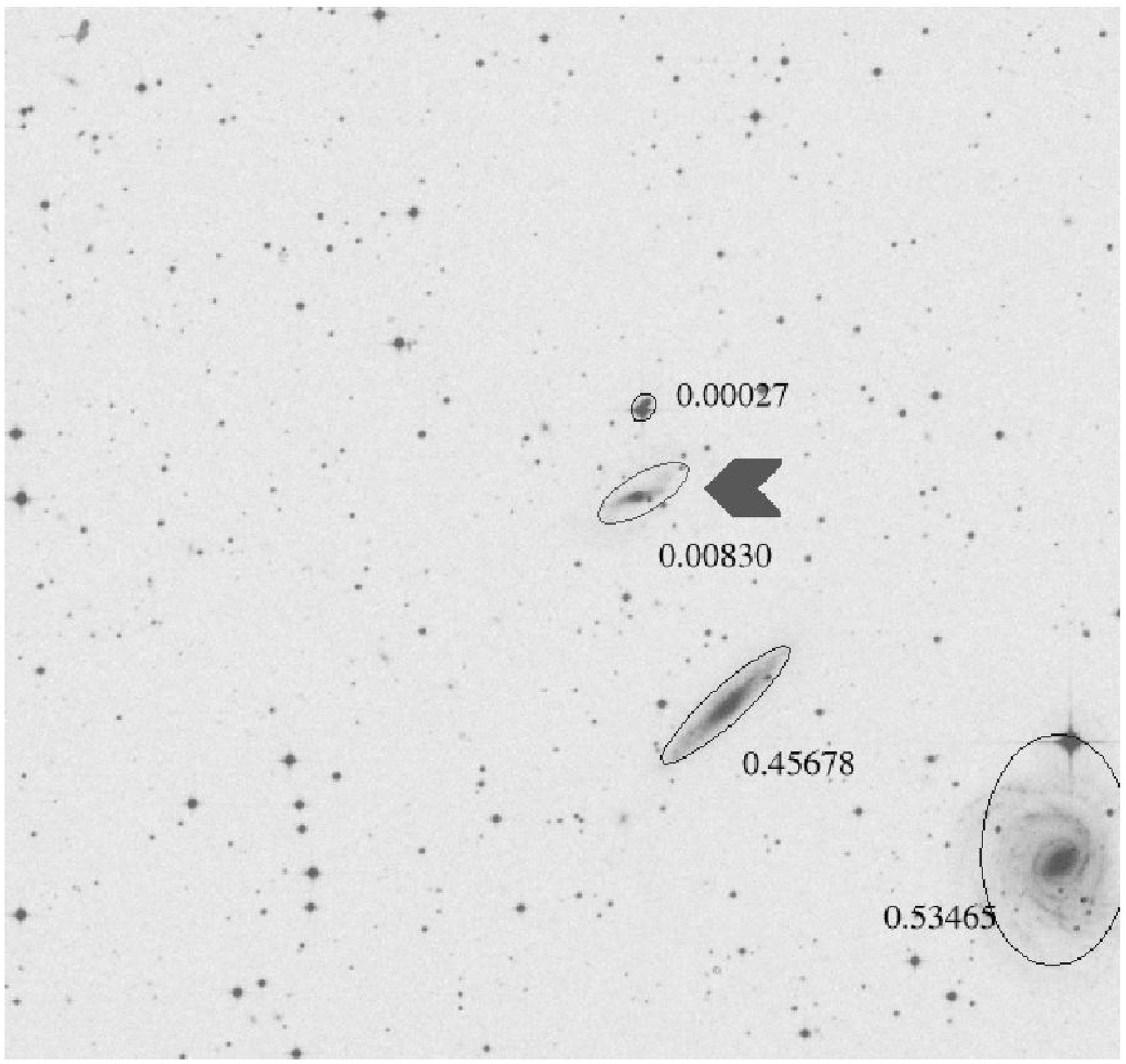}
    \epsfxsize=3.0in
    \epsffile{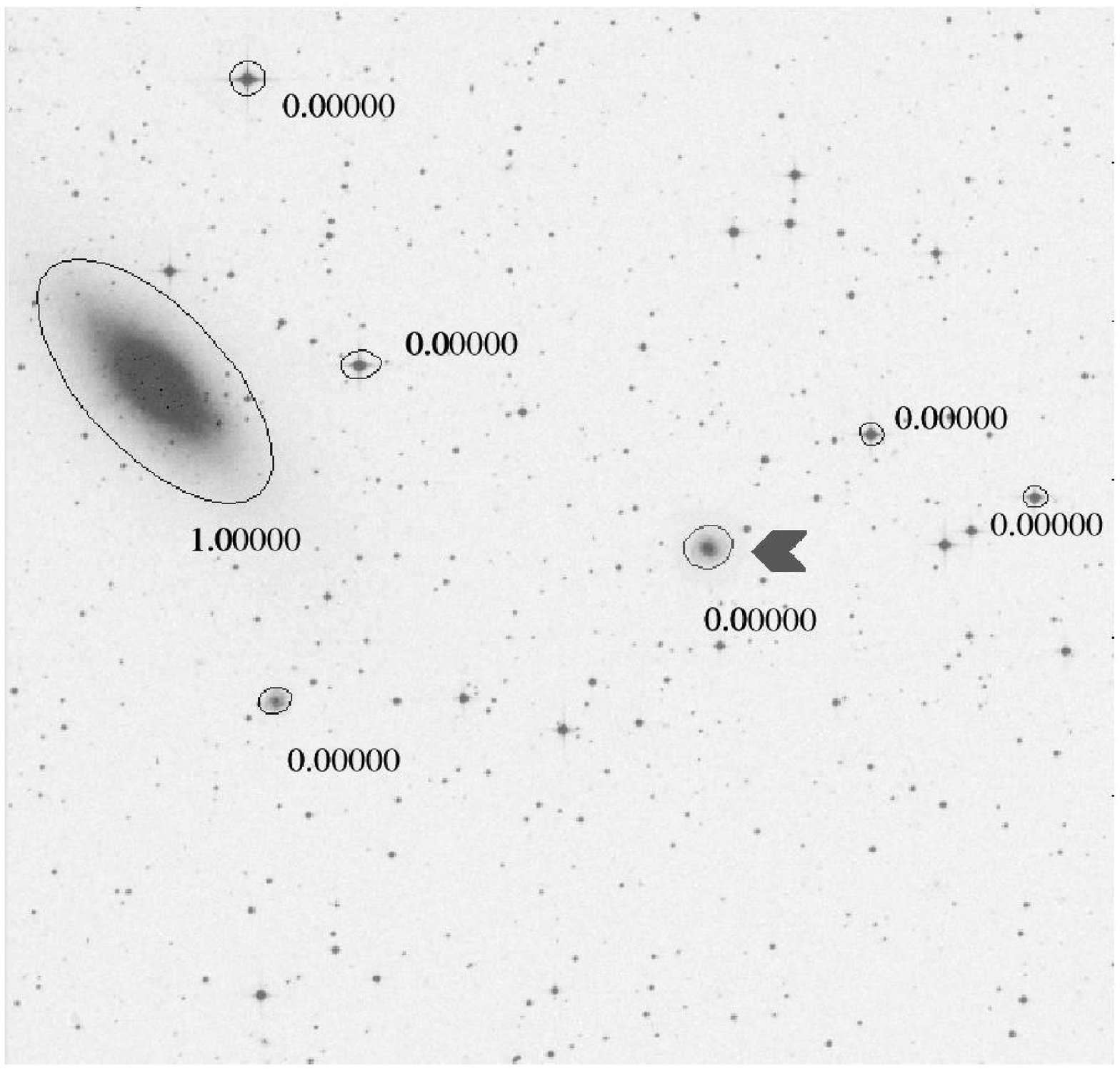}
    }
  }

  \hbox{\hspace{0.5in} (c) \hspace{2.8in} (d) } 
  \vspace{9pt}

  \caption{Some examples of incorrect matches (match marked with an arrow).  
  Two qualitative issues are relevant here: are these difficult cases? 
  Does the classifier 
  predict a reasonable probability to the correct match, and only
  slightly greater for the incorrect match?  In (a) and (b) we get an
  encouraging result where the result is qualified by the fact that
  the highest probability is quite low, and the true link is only
  slightly less likely.  In (c) and (d) performance is
  hampered by the presence of very large and bright background
  objects.  In (c) the highest probability
  is quite low, but disappointingly we obtain a low probability for the
  true match.  In (d) within the precision of our calculations we are
  `certain' about an incorrect link.}

\label{incorrectexamples}
\end{figure}

While the empirical validation of our model is helpful, it is also
useful to include examples so that we can see if the predictions agree
with intuition.  In order to do this we use the discriminative SVM to calculate
probabilities for a number of galaxies in order to make a qualitative assessment.
It has already been shown that the Support Vector
Machine can separate the two classes with high accuracy\citep{rohde2} (for those cases
that were not discarded), 
so we are not interested in determining the correct match per se.  Rather we
are interested in illustrating how the combination of candidate
information changes our probabilities.  We can see this in Fig~\ref{candinfo}
where in (a) the probabilities are produced in ignorance of candidate
information these probabilities are not well behaved in that they
do not sum to one.  In Fig~\ref{candinfo} (b) candidate information has been
incorporated.  The galaxy that was initially thought to be a
$77$ per cent chance of a match is revised to a mere $1.1$ per cent
when all candidate information was incorporated.

Another point that we would like to show qualitatively is that the
class predictions are appropriately \emph{confident} on easy examples and
\emph{cautious} on hard cases.  In Fig~\ref{correctexamples} we show a
number of successful classifications, sometimes in difficult circumstances.
Appropriately the classifier quantifies difficult cases using low
probabilities.  It is particularly impressive that the classifier
distinguishes Fig~\ref{correctexamples} (a) and (b) as these correspond
to two different HI sources in very close proximity.  It is pleasing to
see the probabilities are relatively low (less than $70$ per cent).  In
Fig~\ref{correctexamples} (c) and (d) it is encouraging that the
classifier assigns some probability to the background galaxies, but
is able to place the bulk of the probability on the correct label.

In Fig~\ref{incorrectexamples} we examine cases where the classifier
has assigned the highest probability to an incorrect match.  In  Fig~\ref{incorrectexamples} (a) and (b), while
incorrect, the classifier's prediction seems reasonable - the
probabilities are low indicating a high degree of uncertainty.  
Qualitatively, there appears good reason to be
unsure, we should expect that some of the time the match is not the 
highest probability.  In Fig~\ref{incorrectexamples} (c) the classifier is
confused by the presence of two bright objects that we know from
optical spectroscopy are unrelated to the HI source.  This problem is
even worse in Fig~\ref{incorrectexamples} (d) where the classifier is
apparently certain about an incorrect match.   Effectively the
classifier has told us that an event we know occurred, cannot possibly occur!
The reason for this, is that we had a very extreme case present in the
test data for which there is 
no similar case in the training data, the background galaxy in (d) is
a very large and significant galaxy.  A better situation would be for
the classifier to assign some probability to this event.
In near separable data, such as this, calibration is a difficult
problem as little data lies in the unsure region where probabilities
are between $0$ and $1$.

\subsection{Dark Galaxy Candidate Search}

One of many goals of the HIPASS survey is to identify low surface
brightness (LSB) and dark galaxies, that is HI sources without any
(visible) optical counterpart.  This was carried out using the subset
of radio galaxies matched by \citet{doyle} to SuperCOSMOS known as the HOPCAT catalogue.  
The HOPCAT catalogue contained a number
of HIPASS detections where there are no optical counterparts in the
field, however it was found that the absence of any optical sources could
be satisfactorily explained by dust or stars obscuring the view or by a
false HIPASS detection.  In all other cases there was at least one
possible match, so it was concluded that there was no evidence in
HOPCAT for the detection of dark galaxies \citep{doyle}.
                                                                                
The failure of HOPCAT to detect isolated dark galaxies might be
attributed to the large positional uncertainty of HIPASS together with
the relative
depth of SuperCOSMOS, leading to a very low probability of identifying
\emph{isolated} dark galaxies.  Dark galaxies that happen to lie near a
background optical object cannot be identified using HIPASS, however
follow up observation using high resolution radio HI may be able to
identify these objects.  This telescope time is precious so it is a
worthwhile application to use the probability of no optical
counterpart as predicted by our classifier as a method for selecting targets.
                                                                                
We primarily use the same criteria as HOPCAT for the consideration of HIPASS
sources.  There must be an extinction in the $B_j$ passband $< 1$ mag and the object
must not be on the galactic plane or obscured by stars.  However we relax
the assumption that there must be no galaxies in the field to simply
include fields where the probability of no match is high.  This procedure
selects fields where there are few candidates and these objects are more
consistent with being background objects than HIPASS counterparts.

In order to detect dark galaxy candidates we must set the
identification rate $\kappa$ to greater than $0$; we rather arbitrarily set
$\kappa = 1.6 \times 10^{-4}$ or
$\kappa\frac{P(z_{i,j}=1)}{P(z_{i,j}=0)} = 0.001$.  The choice of
$\kappa$ is very subjective, it depends on a priori belief in the
existence of dark galaxies.  However regardless of the choice of
$\kappa$ the output probability is a monotonic function of the
probability of a dark galaxy, so this is no obstacle to creating a
list of best candidates for follow up observation.

When $\kappa > 0$ the probabilities for each candidate in a field will not sum to 
one.  The probability of no match (dark galaxy) is one less the sum of all
candidate probabilities.  The HI sources that are most likely to be
dark galaxies will then be those that have few good candidate
matches.  Using this procedure we produced a list of dark
galaxy candidates and the probability that they have no match in Table 1.

The list contains the $30$ objects with highest probability of being a
dark galaxy according to our classifier.  They also satisfy the
criteria of 
having the extinction in the $B_j$ passband $< 1$ mag  and have not previously been eliminated as
candidates by \citet{doyle}.  The description column contains
comments obtained from doing a visual inspection of each field.  Fields
containing small galaxies would be particularly interesting to follow
up with high resolution HI observations.  Follow up observations could help
to determine (i) if there are more false HIPASS detections, (ii) locate
dark galaxies or (iii) identify HIPASS sources that are
very small or faint in SuperCOSMOS.  It would also have implications for the
reliability of the HOPCAT catalogue.

\begin{figure}

  \centerline{\hbox{ \hspace{0.20in} 
\label{roc}
    \epsfxsize=3.0in
    \epsffile{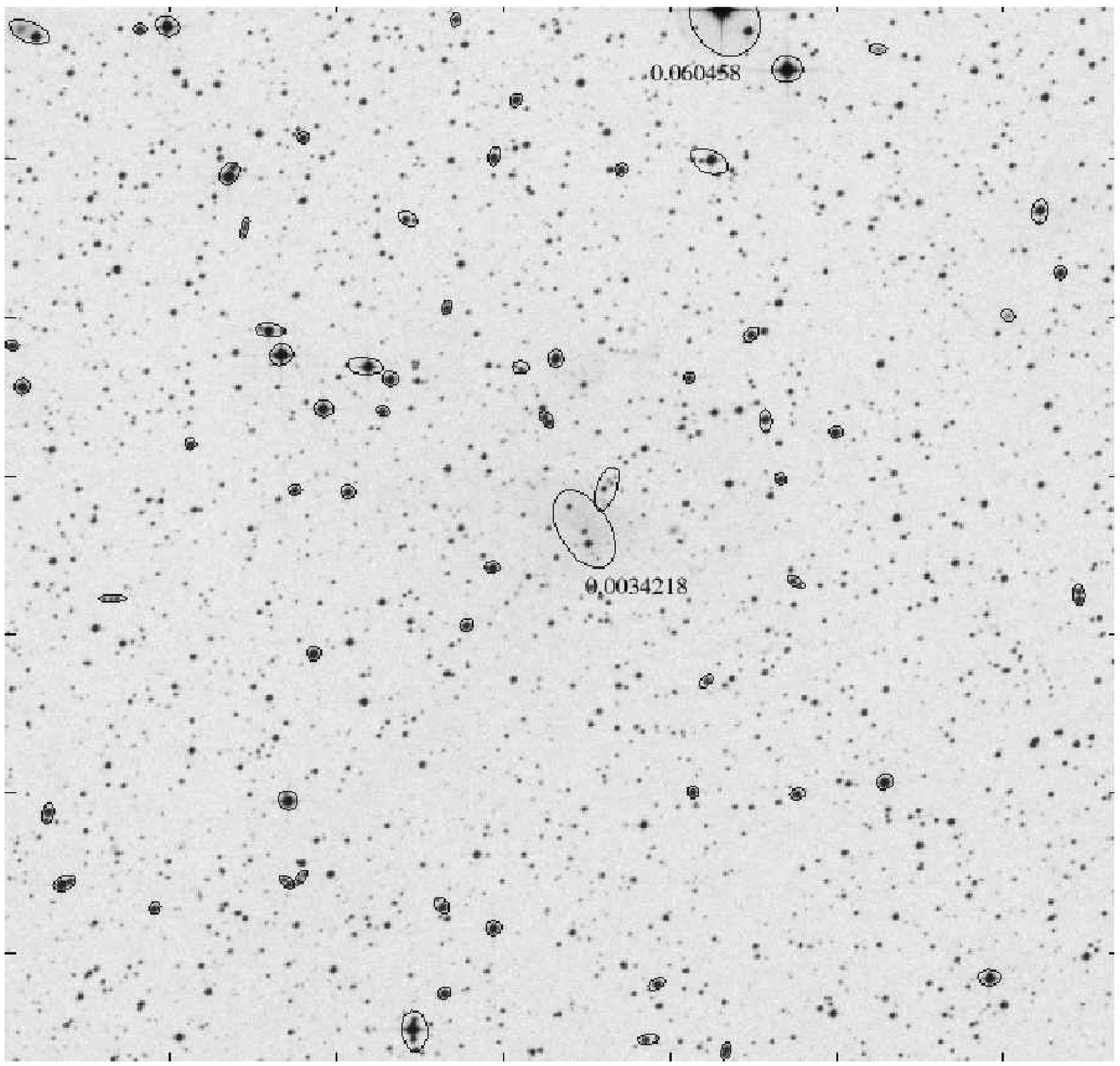}
    \epsfxsize=3.0in
    \epsffile{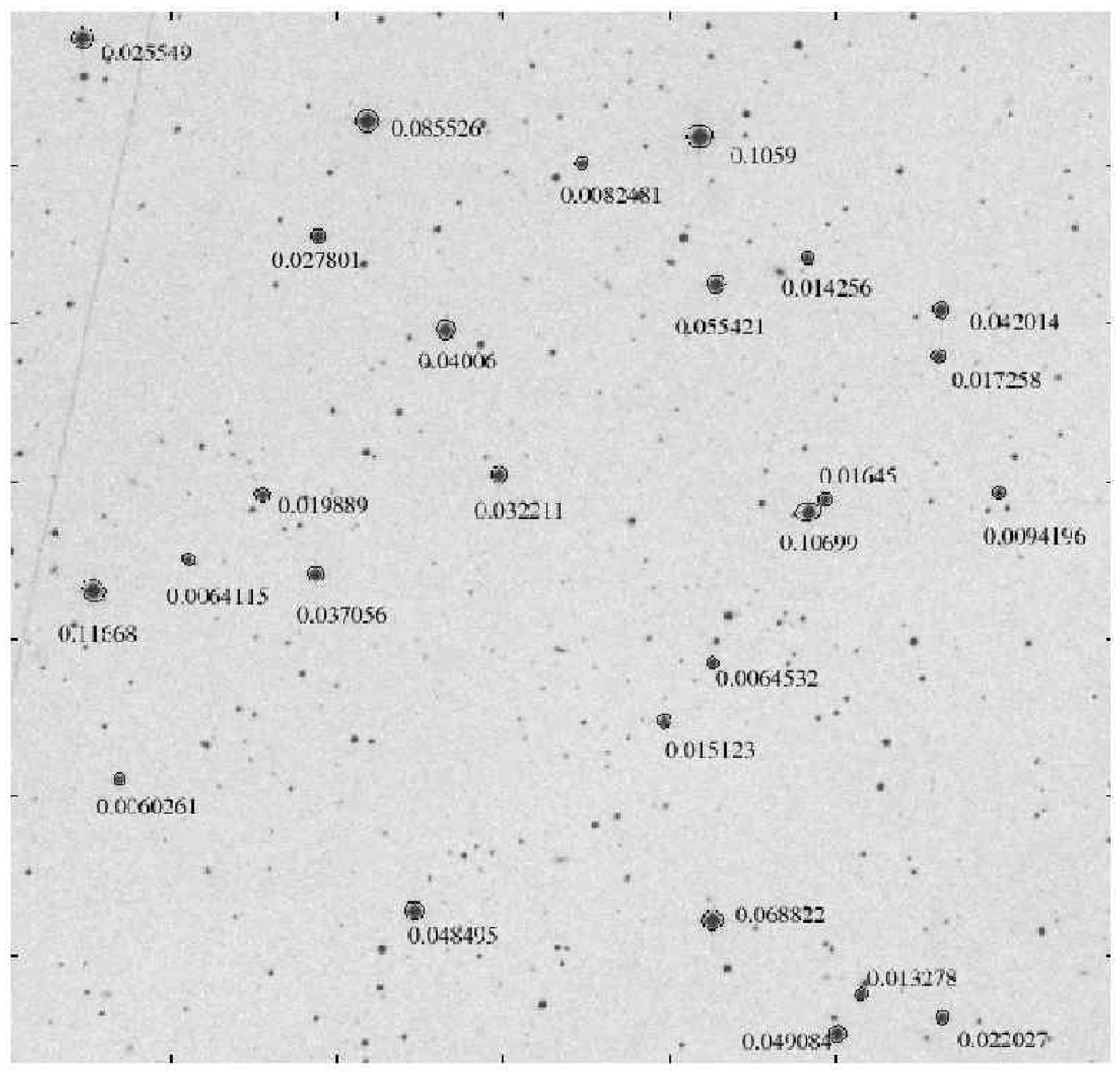}
    }
  }

  \vspace{9pt}
  \hbox{\hspace{0.5in} (a) \hspace{2.8in} (b) } 
  \vspace{9pt}
\caption{Two examples of the probabilities of candidates around
dark galaxy candidates.  In (a) J105732-48 which is classed as a good
guess in HOPCAT - viewing the image in ds9 and varying contrast makes
it seem credible that the central object is a match.  It is
interesting nonetheless as the galaxy is quite diffuse.  Probabilities
that are less than $0.0005$ are not shown.  In (b) J0532-78
no strong candidate counterpart exists.  Relatively high probability
is assigned to the stars, partially reflecting the inability of the
classifier to identify stars with absolute confidence.  However it is
an affect of the absence of any reasonable candidate causing
substantial probability mass to be assigned to the stars.
Intuitively, when we
further condition on our own belief that these stars cannot be
matches to the HI source the probability that this object is rare
increases to be greater than the $0.01$ as previously calculated.}
\label{darkgalimage}
\label{roc}
\end{figure}

Two fields containing examples of dark galaxy candidates are shown in Fig~\ref{darkgalimage}.  The images are most strongly characterised by
the lack of large and significant galaxies.  The description column of
Table~\ref{darkgal} indicates this is typical; the fields containing
larger galaxies may have optical properties that indicate they are
unlikely to contain significant amounts of HI or perhaps, more likely,
are cases where the classifier performs badly.

\begin{table}
\begin{tabular}{|l|r|l|l|r|r|}
\hline
Name&Probability&Description&Class&Sint(Jy km $s^{-1}$)&Velocity (km s$^{-1}$)\\
\hline
J1057-48&0.94&LSB candidate match in centre&gg&101.7&597.6\\
J2350-40&0.58&Near empty field&no-vel&33.5&1698.4\\
J2355-39&0.47&Several faint galaxy&no-vel&21&263.4\\
J0033-09&0.27&Near empty field&no-vel&4.6&2751.5\\
J2251-20&0.15&Two good candidate matches&no-vel&14.8&3166.7\\
J1341-02&0.12&Near Empty Field&no-vel&3&8820.9\\
J2250+00&0.11&LSB galaxy in centre&no-vel&2.8&1696.5\\
J2351-40&0.11&Several galaxies near centre&no-vel&7.6&344.3\\
J1225-06&0.10&Some galaxies near centre&no-vel&6.2&1231.4\\
J0623-42&0.09&Galaxy near centre; crowded with stars!&no-vel&10.7&2259.3\\
J1435-17&0.07&LSB at high separation&no-vel&9.1&1576\\
J1227+01&0.05&LSB near centre&no-vel&9.1&1576\\
J1024-12&0.05&LSB near centre&no-vel&31.9&1292.4\\
J0951+01&0.04&LSB near centre&no-vel&2.6&623.3\\
J0249+01&0.03&LSB near centre&no-vel&2.4&2936\\
J1045-83&0.03&Spiral near centre; crowded field&no-vel&9.73&2123.2\\
J0214-13&0.02&Good candidate match near centre&no-vel&3.2&5812.3\\
J1438-18&0.02&Good guess near edge&no-vel&12.3&2562.7\\
J0013-26&0.02&Near empty field&no-vel&3.7&4871.6\\
J2207-75&0.02&LSB candidate near centre&no-vel&2.9&2796.9\\
J2150-23&0.02&Faint galaxies&no-vel&5.2&2346.5\\
J1334-12&0.02&Faint galaxies&no-vel&2.2&1503.7\\
J0958-85&0.02&Faint galaxies; crowded with stars&no-vel&2.1&1976.1\\
J2331+01&0.02&Two reasonable candidate matches&no-vel&7&1271.5\\
J1347-30&0.01&Several reasonable candidate matches&no-vel&53.1&4358.3\\
J1812-74&0.01&Faint galaxies; crowded with stars&no-vel&3.8&3199l.7\\
J0946-74&0.01&Good candidate match; crowded with stars&no-vel&46.9&1152.9\\
J0648-84&0.01&LSB good candidate match&no-vel&3.5&5287.9\\
J0909-83&0.01&LSB good candidate match&no-vel&3.9&2033.1\\
J0532-78&0.01&Some faint galaxies; crowded field&no-vel&4.1&6103\\
\hline
\end{tabular}
\caption{Dark Galaxy Candidates.  The probability column refers to the
  predicted probability that this HI source has no counterparts and
  the description column is from
  visual inspection.  As explained in the text the interpretation of
  the probability is dependent on a subjectively determined $\kappa$
  which is higher for higher beliefs in dark galaxies.
  The abbreviation LSB is used for low surface 
  brightness galaxies.  HOPCAT Class describes how class was determined
  in HOPCAT; gg means the object was a good guess i.e. it was judged
  that there was only one likely candidate; no-vel means that many
  galaxies are present in the image, but redshift information is
  insufficient to determine a match.  The final columns are the integrated
  flux measured from HIPASS, and the Velocity from HIPASS.  The
  average flux of HIPASS is 15.76 Jy km s$^{-1}$ and the mean velocity is $3275.14$ km s$^{-1}$ over the entire sample.}
\label{darkgal}
\end{table}

\section{Biasing Limitation}
\label{future}

Ideally the matched output would recover the underlying distribution
$P(\alpha, \beta|z=1)$
in the output it produces, however we note here that biasing effects
are possible.

A development in this paper is the use of probabilities conditioned on
all available data.  It is tempting to pair every sparse object with the most likely dense candidate and then to act as if this combined catalogue is completely true in the analysis to follow.  This has the advantage that we find
the most likely match for \emph{every} sparse object.  In contrast the
binary SVM produced $1012$ ambiguous results that had to be
discarded.  However we note here that keeping all information comes at
a price: choosing the most likely candidate makes the output
distributions distorted.  We illustrate this effect with a simplified example.

\begin{figure}
\center
\includegraphics[width=3.5in]{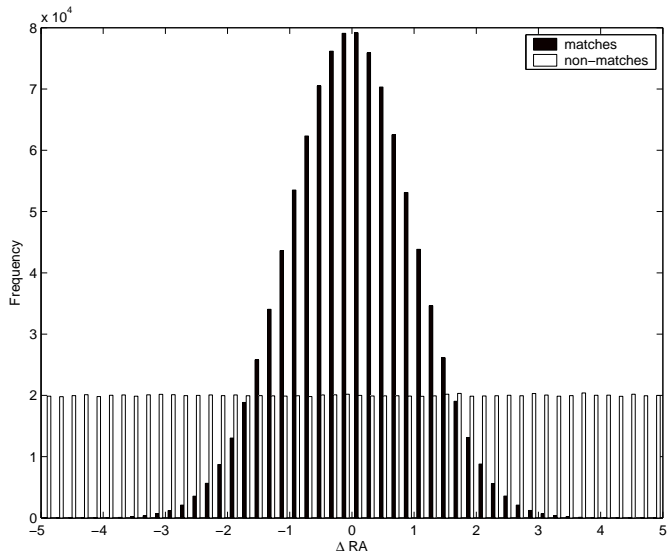}
\caption{Our problem is to determine which objects match given
  position.  From this histogram it is clearly impossible to establish
  this with certainty, and that the optimal procedure given any pair
  of links is to pick the one with the smallest proximity.}
\label{mismatch}
\label{problem}
\end{figure}

Consider the simplest possible problem where we match objects using a
single parameter $\Delta RA$.  We only consider objects with a
$\Delta RA$ less than $5$ arcsec.  The distribution for $\Delta RA$ for matched
objects has a Gaussian distribution of mean $0$ arcsec and standard
deviation of $1$ arcsec and
the non-matching objects are uniform between $-5$ and $5$ arcsec.  For each
field there is exactly $1$ match and exactly $1$ non-match.  The
distribution of matches and non-matches is represented in Fig~\ref{problem}; the overlap shown is the
Bayes risk i.e. the unavoidable error rate when classifying.

The analogue of the binary classification approach in 
this situation is to accept a match if, and only if, there is exactly
one object in the accept region (for the purposes of this discussion
we will accept matches between $-3$ and $3$ arcsec).  If there are multiple or
zero objects in this region then the data is discarded.  The analogue
of the probabilistic method discussed here is to take the most
probable object as being the match (and act as if this is completely true); no data is ever discarded.

\begin{figure}
  \centerline{\hbox{ \hspace{0.20in} 
    \epsfxsize=2.0in
    \epsffile{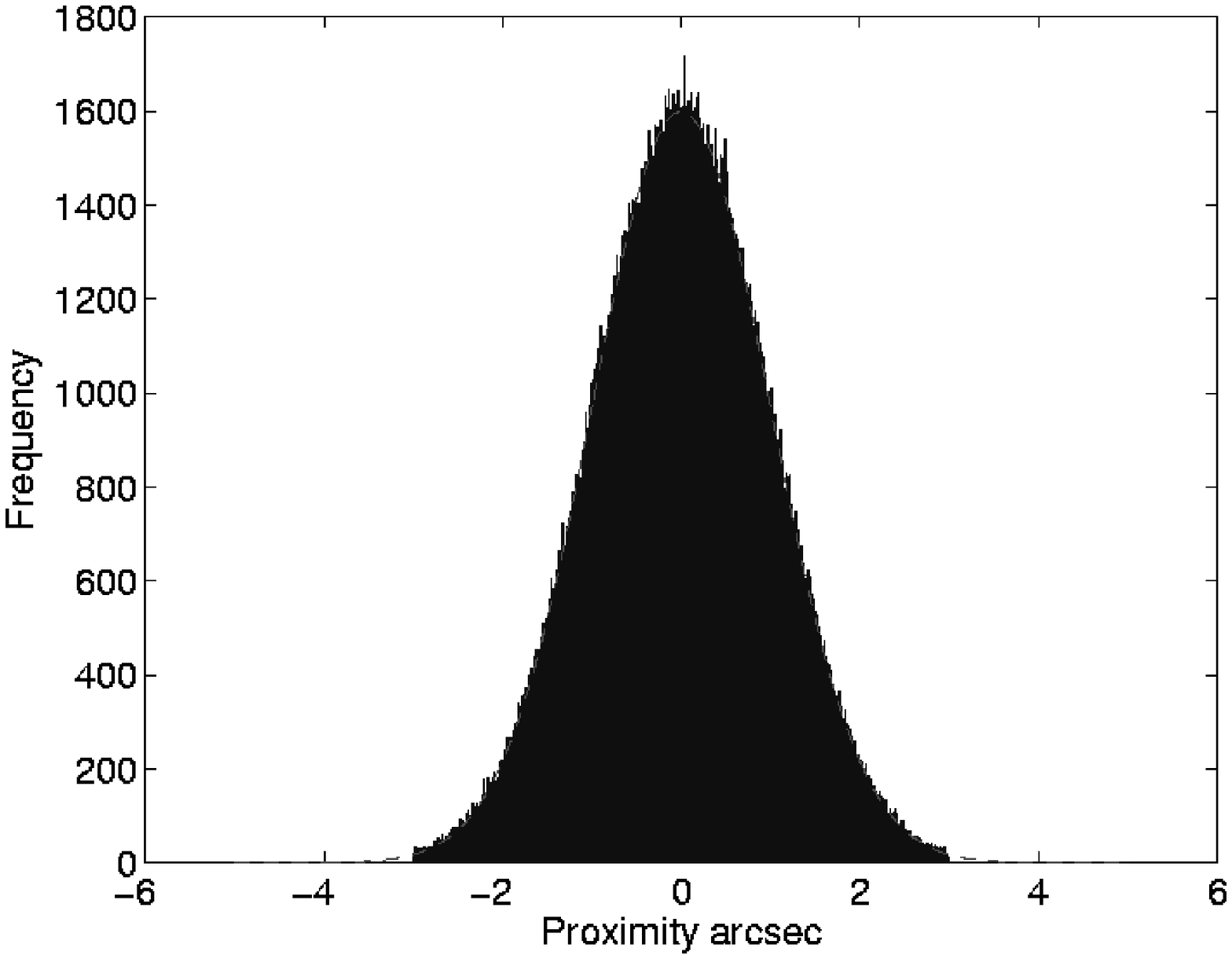}
    \epsfxsize=2.0in
    \epsffile{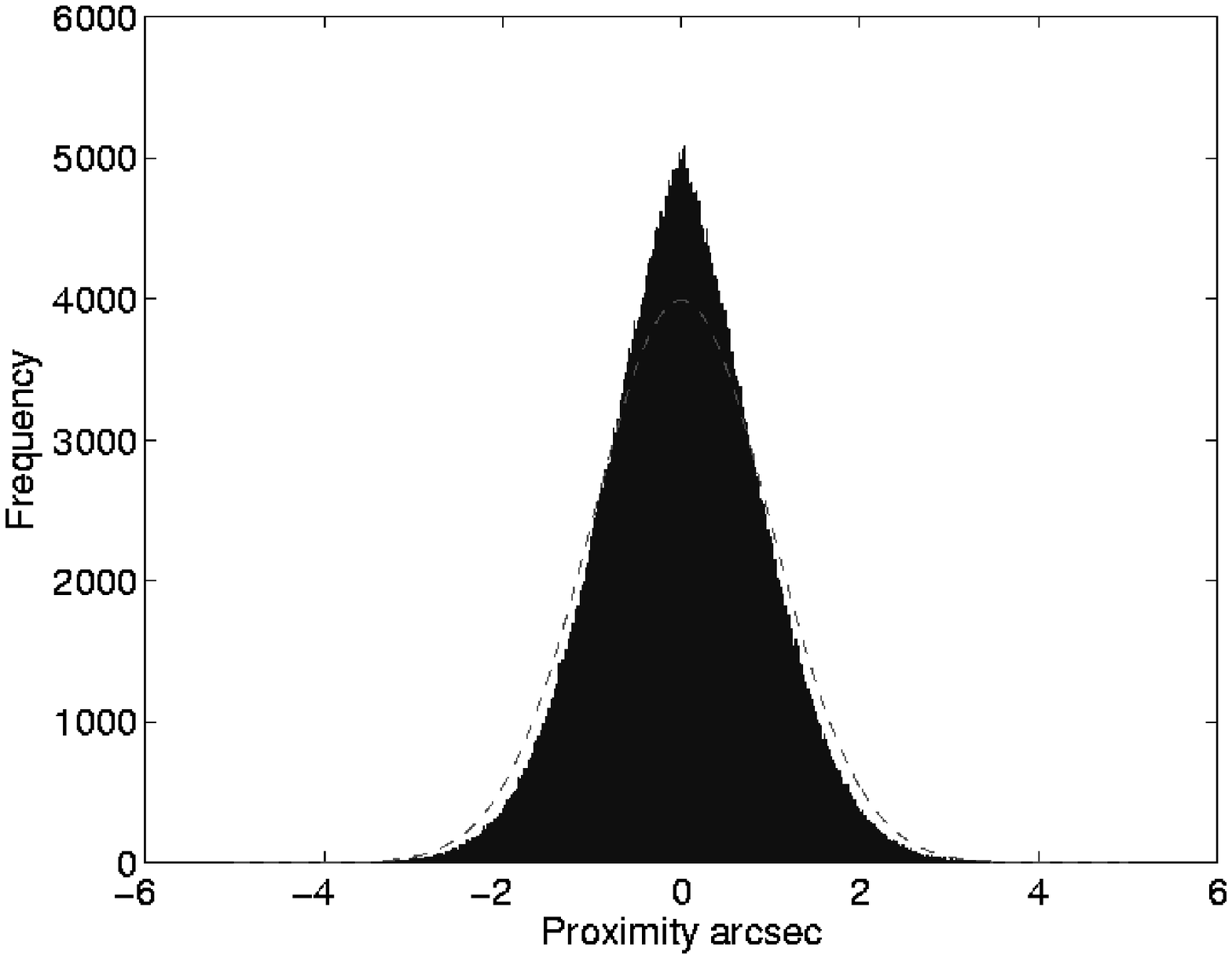}
    \epsfxsize=2.0in
    \epsffile{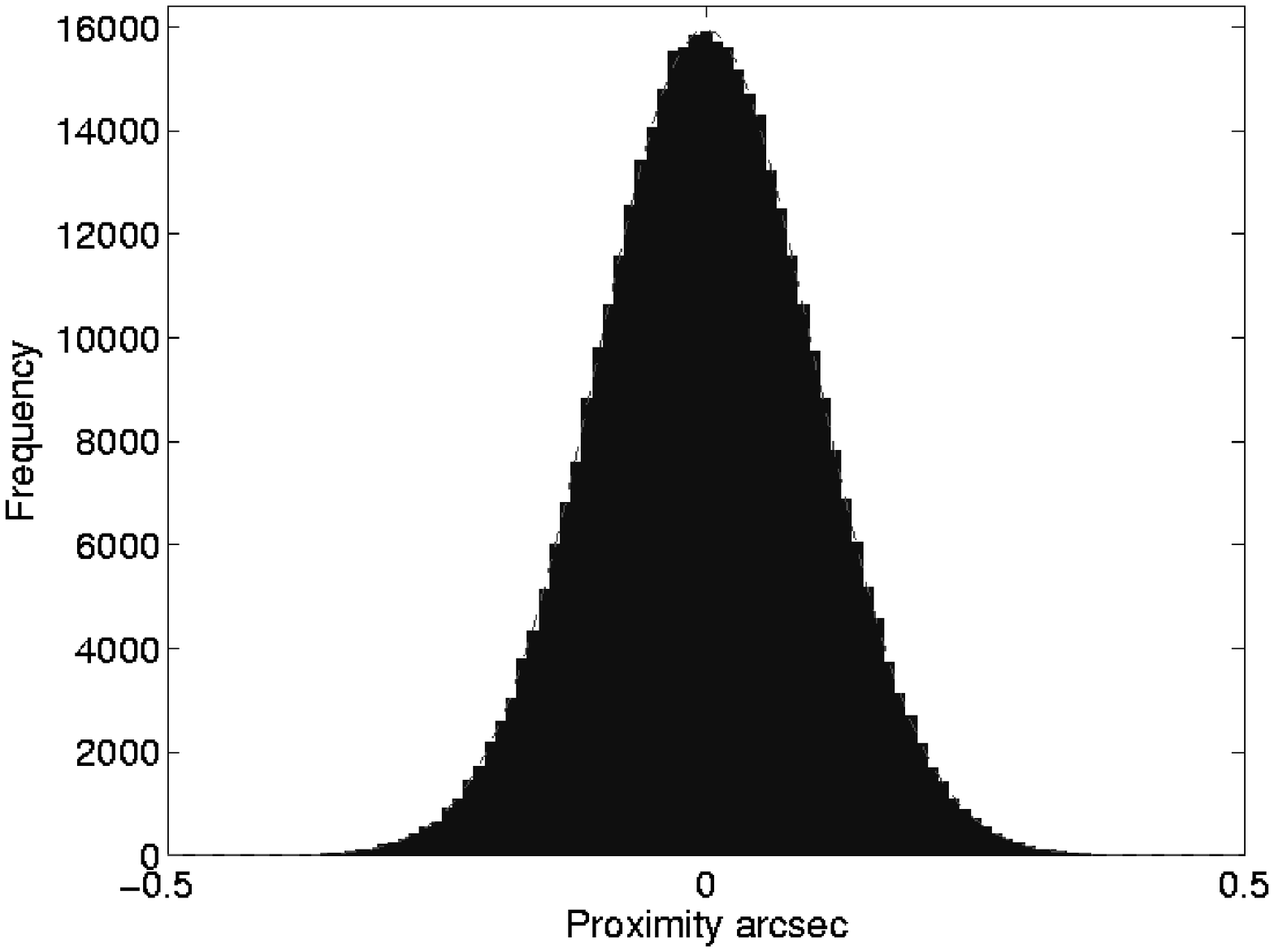}
    }
  }

  \vspace{9pt}
  \hbox{\hspace{0.6in} (a) \hspace{1.8in} (b) \hspace{1.9in} (c)} 
  \vspace{9pt}

  \centerline{\hbox{ \hspace{0.20in}
    \epsfxsize=2.0in
    \epsffile{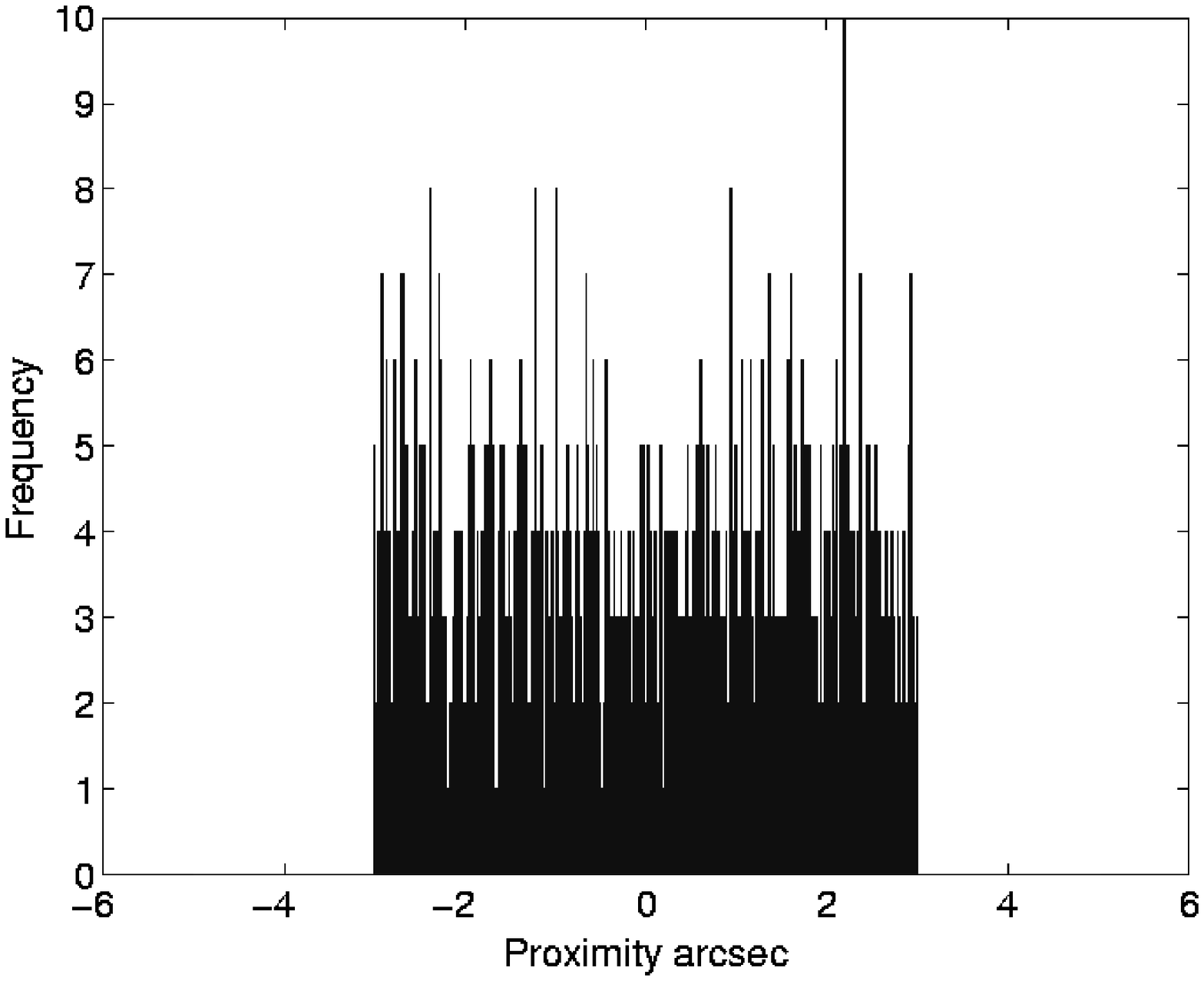}
    \epsfxsize=2.0in
    \epsffile{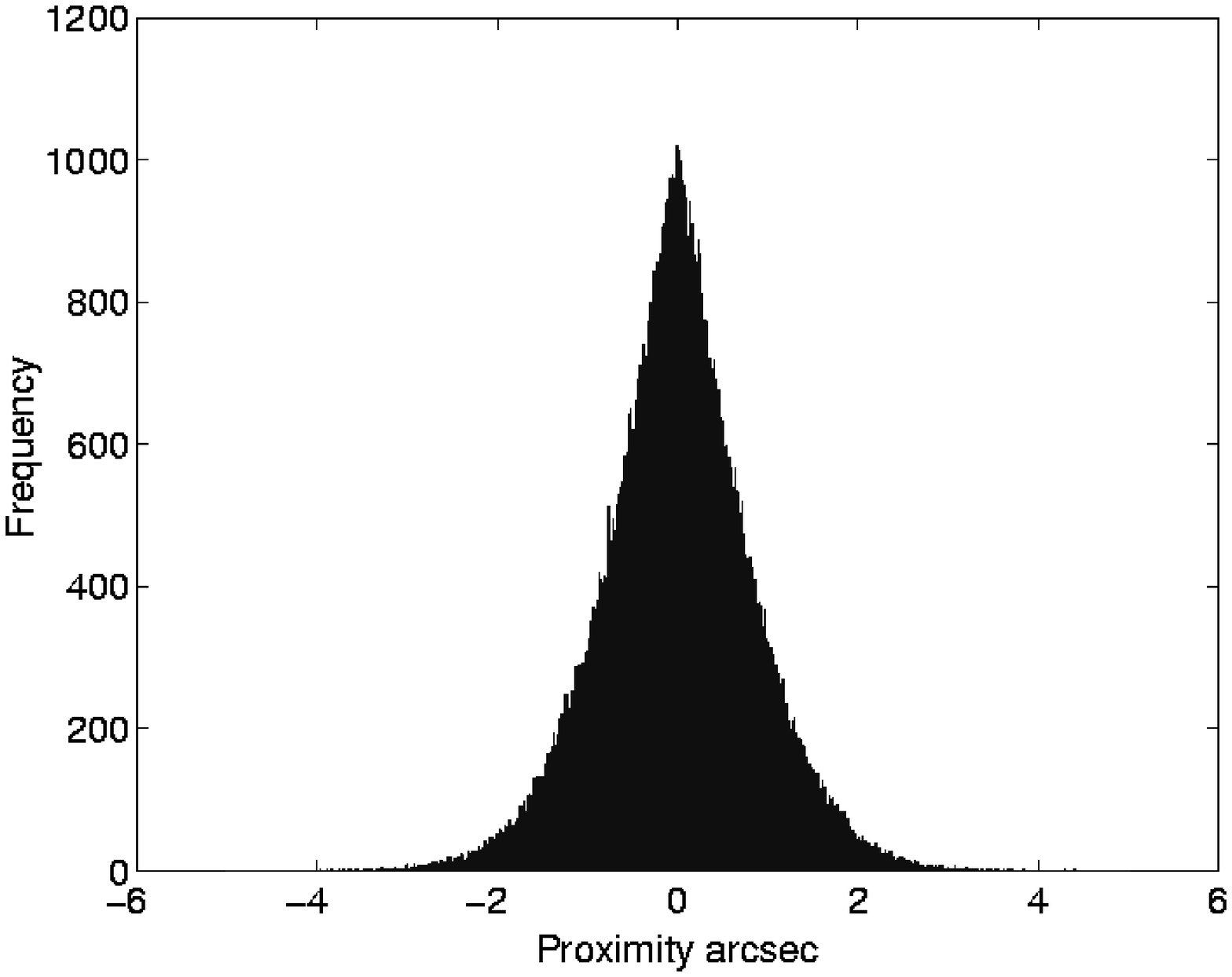}
    \epsfxsize=2.0in
    \epsffile{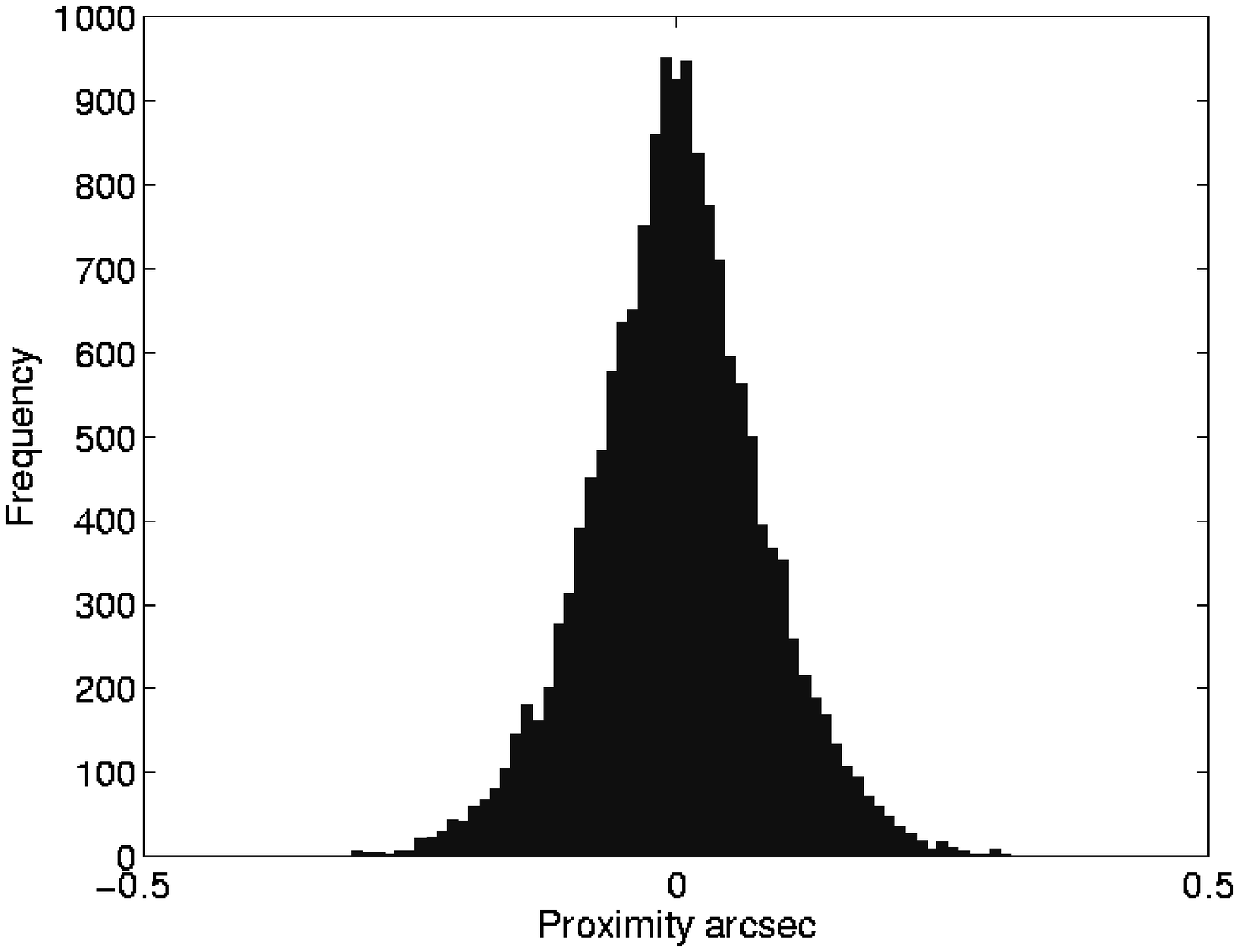}
    }
  }

  \vspace{9pt}
  \hbox{\hspace{0.6in} (d) \hspace{1.8in} (e) \hspace{1.9in} (f)} 
  \vspace{9pt}

 \centerline{\hbox{ \hspace{0.20in}
    \epsfxsize=2.0in
    \epsffile{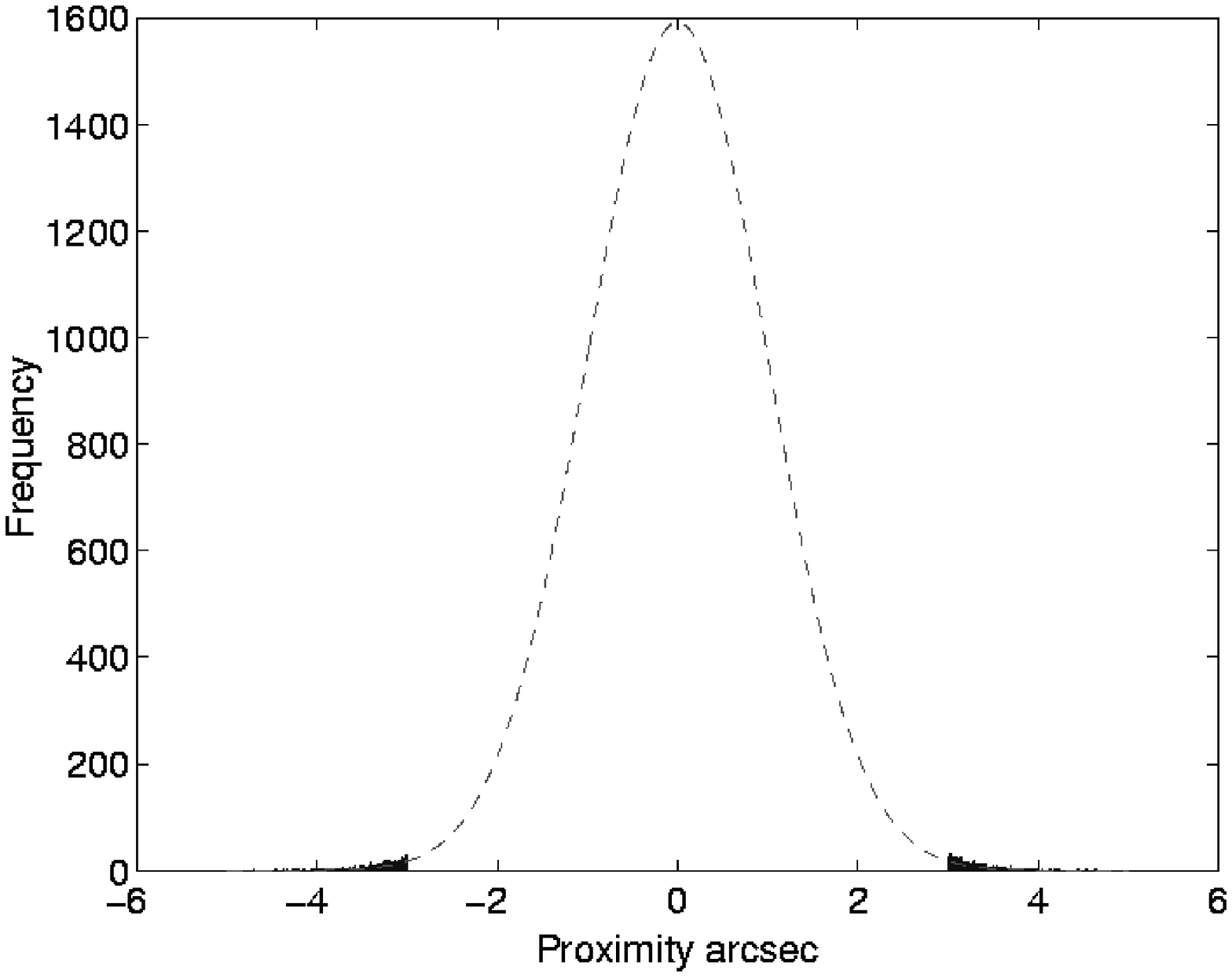}
    \epsfxsize=2.0in
    \epsffile{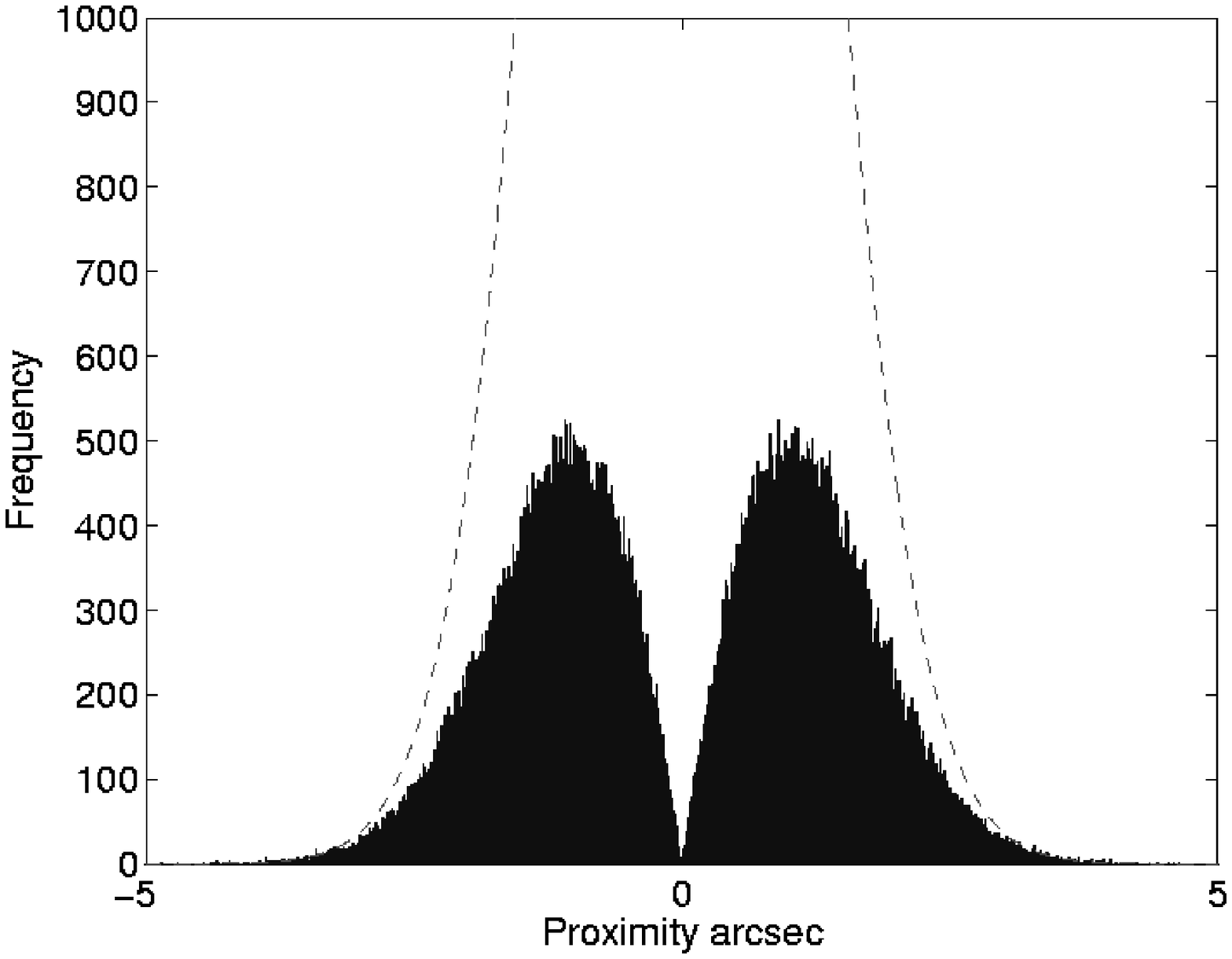}
    \epsfxsize=2.0in
    \epsffile{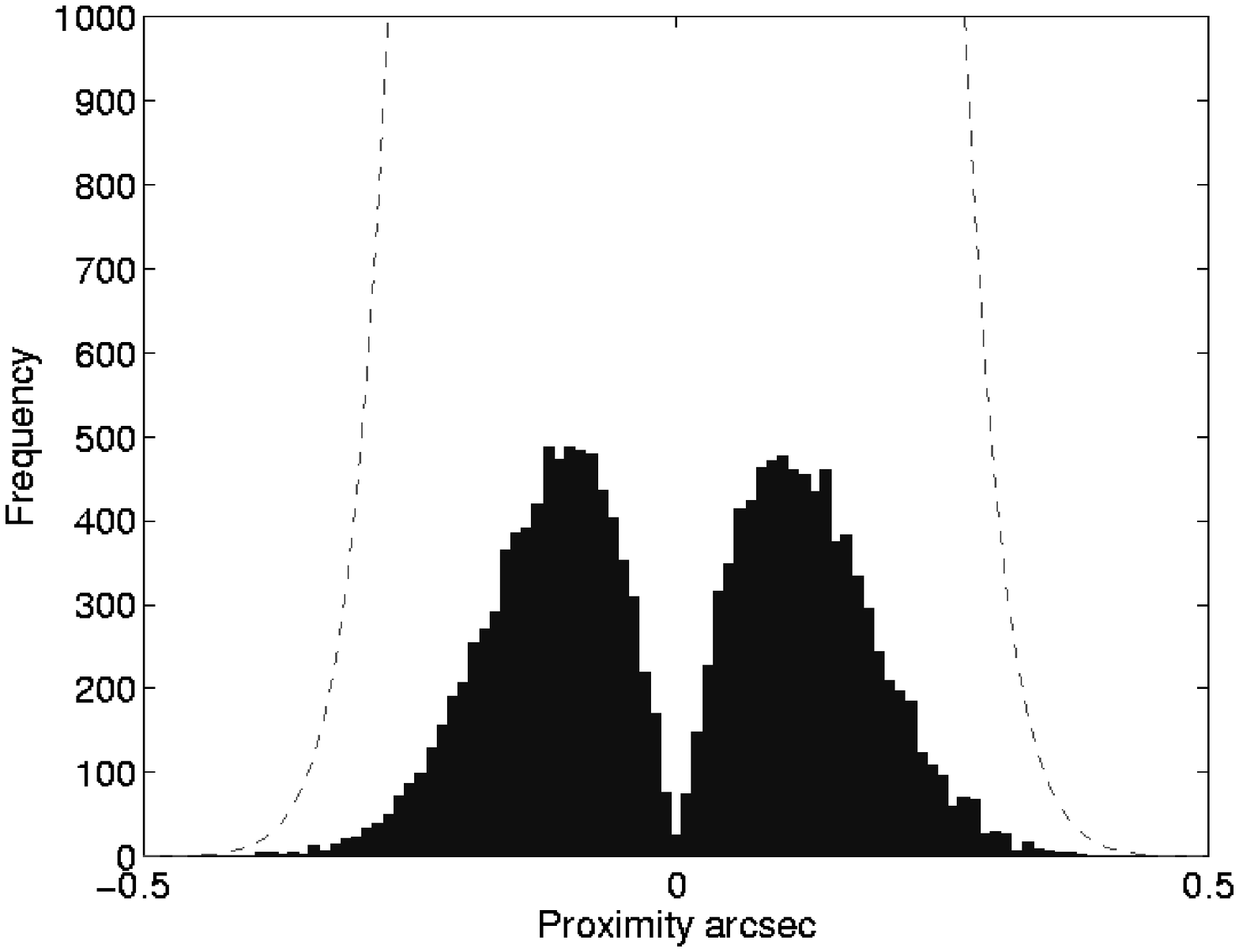}
    }
  }

  \vspace{9pt}
  \hbox{\hspace{0.6in} (g) \hspace{1.8in} (h) \hspace{1.9in} (i)} 
  \vspace{9pt}

  \caption{The distributions recovered from our simple matching
  simulation.  The thresholding output is shown in (a) the false
  positives in (d) and the false negatives in (g) we note that the output
  distribution is a good recovery of the Gaussian distribution (dashed
  line) the cost of this is that $30$ per cent of data was discarded which is
  a significant waste of information.  The output by selecting the
  most likely match is shown in (b) this is noticeably different from
  Gaussian - it is a tighter distribution, the false positives are
  shown in (e) and the false negatives in (h).  These simulations dwell
  on a pessimistic case, if we attempt an easier problem where the
  standard deviation of the Gaussian is reduced from $1$ arcsec to
  $0.1$ arcsec the
  error for the most likely match procedure drops to $1.6$ per cent.
  The output is then satisfactorily close to Gaussian (c) and the
  false positives (f) and false negatives (i) are (perhaps) negligible.}
  \label{misclassifications}
\end{figure}

We generate $10^6$ random classification problems.  This is done by
drawing one sample from the Gaussian and one from the uniform
distribution.  The two different approaches are applied, the most
likely match always assigns a class, the binary classifier will either
assign a class or discard the data.  We then consider the output
distribution for the false positive distribution and the false
negative distribution.  Fig ~\ref{misclassifications} (a)
shows the recovered output using the thresholding decision rule, the
output is near Gaussian but has the tails cut off 
 at $<-3$ and $>3$ arcsec.  Also $30$ per cent of data was discarded.
 Of the data that was retained only $0.4$ per cent was erroneous: the
 false negatives (Fig ~\ref{misclassifications} (d)) are the lost tails and, the
false positives represent a uniform distribution which has been added
to our recovered output.  This is a relatively simple situation with a
near Gaussian distribution recovered at the cost of a lot of data
being discarded.

When selecting the most probable match no data is discarded, however
the error rate increases to $16$ per cent.  Moreover the distribution of
errors appears more complex.  The output
obtained is not Gaussian (Fig ~\ref{misclassifications} (b)).  The false
positives are clustered around low proximity (Fig
~\ref{misclassifications} (e)); and the false negatives (Fig
~\ref{misclassifications} (h)) show a bi-modal distribution representing
unusually high proximity objects that were discarded exactly for that
reason: they are not typical matches.  This has interesting
implications for using catalogue matching to search for rare objects
which is seen as one of the avenues for new science in the VO
\citep{djorgovski} and in which the ClassX work is starting to make some
progress \citep{2004ApJ...612..437S}.

This example just discussed exaggerates the biasing effects, at least
with respect to our HIPASS-SuperCOSMOS dataset.  The
error rate of $16$ per cent in comparison to the problem used in this
study is unrealistically high.  The error rate in this simulation is
determined by the variance of the Gaussian and the
threshold where objects are not included (beyond $\pm5$).  If we
alter this by reducing the standard deviation to $0.1$ then the error
rate becomes $1.6$ per cent and we find 
that the recovered distribution (Fig ~\ref{misclassifications} (c)) is
near Gaussian and we can more or less disregard false positives (Fig
~\ref{misclassifications} (f)) and false negatives (Fig
~\ref{misclassifications} (i)), note the change of scale on the x-axes
in these figures.

A reasonable amount of knowledge of the underlying distributions is
required in order to match objects together.  In the approach taken
here we actually choose to act as if our density estimates are
completely true (of course they are not).  In this situation where we
already have a great deal of information what do we gain by using our newly
matched data?  Why not simply use the linked 
subset for inference and skip the complexity of this matching step?
Precisely how much information do we lose by ignoring some of our 
data?  A possible way to consider this problem is to use information theoretic
concepts such as \emph{information gain}.  In this situation the problem of
scientific inference is seen as equivalent to constructing a
communication channel which can transmit the scientific `truth' using
the smallest number of bits.  Constructing a communication channel using
probabilities from the matched subset is likely to work reasonably
well.  However a more efficient communication channel would also employ
the hard-won information from the newly matched data.  The difference
in the efficiency of the two communication channels has intuitive
appeal as the information gain.  In a Bayesian
approach this additional information is calculated as the
Kullback-Leibler divergence between the prior and the posterior.  This
is a common heuristic for measuring the expected information gain see
\citep{lindley_info, bernardo_info}.  In our application if the expected
information gain is low it seems reasonable that the considerable
effort to gain information from catalogue matching may not be
worthwhile.  In future work we plan to investigate this further.

\section{Summary}

In this paper we presented a way to view the statistical problem of
catalogue matching or linkage in terms of a combination of either
generative pdf estimators or discriminative pattern classifiers.  The
discriminative method however fails to include a very reasonable
independence assumption.  We
applied each approach to the problem of matching HIPASS-SuperCOSMOS
and showed two approaches to using all available information.  While
fitting these models we were unable to provide an absolute criterion for
an optimal probabilistic estimate, however we provided a number of
heuristics to test if the probabilities are satisfactory.

Probabilities are useful as they allow completeness efficiency
trade-offs to be controlled, they also make it possible to search for
rare objects.  A list of $30$ dark galaxy candidates were provided using
probabilities in order to produce a ranked list.

The study produced here is in some sense easier than many matching
tasks that may be performed in that a training set of matched examples
was already available \footnote{In order to construct a model without
  training data, \cite{sutherland} recommended a procedure that involved
histogram subtraction of two distributions although this has been
found to give unsatisfactorily noisy estimates \citep{mann97}.  For a
more sophisticated way of estimating these densities see \cite{storkey}}.  The
availability of a training set allows the density estimation to be
performed in high dimensional space using semi-parametric models such
as Gaussian Mixture Models.  This allows a direct application of the
\cite{sutherland} formalism which appears to give reasonable results.  From a
machine learning perspective this would be termed a generative model.

On-going debate in machine learning and statistics communities continues
on the relative merits of generative and discriminative methods for
classification.  The discriminative method is often preferred in the
literature, however for our problem it is non-trivial to obtain a
probability conditioned on all available parameters.  It also causes an
independence assumption to be ignored that we know applies a priori.
Despite this, the classification results of a probabilistic
(discriminative) SVM are
very competitive with the generative model.  In terms of
classification the (discriminative) SVM had a slight edge over the
generative Gaussian Mixture Model.

It turns out that, more generally, the question of `is this probability a
good probability' is not well posed.  We offer three heuristics checks,
classification rates, calibration diagrams and Brier scores for evaluating the quality of
the probabilities, but none could be considered definitive.  Both
the SVM and Gaussian Mixture Model performed well against these
intuitive measures; the question is not sufficiently well posed to
make any statement about one being better than the other.

The utility of probabilities was demonstrated in that it was possible
to produce a matched catalogue with some control over completeness and
efficiency.  Moreover probabilities were useful in obtaining a list of
candidate dark galaxies for which follow up observation with the
Australia Telescope Compact Array may be informative.

This paper generalises the framework of \cite{sutherland} to deal with sparse
parameters as well as dense parameters and considers the problem using
high dimensional pdfs.  The problem of estimating the distributions
of interest is open, however we have shown two alternative methods both
can provide good results.  We are considering another approach using
Bayesian inference approximated using Markov Chain Monte Carlo
algorithms for a future paper.

\bibliographystyle{mn2e}
\bibliography{mybib}

\section*{Appendix A - The Sutherland and Saunders Result}

\noindent
Assuming that each $a$ links to at most one $b$ then we can say:

\begin{equation}
P(\alpha_x,\beta_{x,1}, \cdots,\beta_{x,N_x} |z_{x,k}=1) =
  P(\alpha_x,\beta_{x,k}|z_{x,k}=1) \prod_{i=1..N_x,
  i \ne k} P(\alpha_x, \beta_{x,i}|z_{x,j}=0).
\label{indepen}
\end{equation}

\noindent
Similarly if $a$ links to \emph{no} $b$ then we can say:

\begin{equation}
P(\alpha_x,\beta_{x,1}, \cdots,\beta_{N_x} |z_{x,1}=0, \cdots, z_{x,N_x}=0) =
  \prod_{i=1..N_x} P(\alpha_x, \beta_{x,i}|z_{x,i}=0).
\label{indepen2}
\end{equation}

\noindent
The quantity of interest is

\begin{eqnarray}
P(z_{x,y}=1 | \alpha_x,\beta_{x,1}\cdots, \beta_{x,N_x}) = &\nonumber \\
&\frac{P(\alpha_x,\beta_{x,1}, \cdots, \beta_{x,N_x} |z_{x,y}=1)
  P(z_{x,y}=1) }{\sum_{j=1..N_x} P(\alpha_x,\beta_{x,1},
  \cdots,\beta_{x,N_x} |z_{x,j}=1) P(z_{x,j}=1) +
  P(\alpha_x,\beta_{x,1},
  \cdots,\beta_{x,N_x}|z_{x,1}=0, \cdots, z_{x,N_x}=0)P(z_{x,1}=0, \cdots, z_{x,N_x}=0) }
\end{eqnarray}

\noindent
Assuming the priors belief is the same ($\frac{1}{N + \kappa}$) for all
candidates.  It follows that the prior belief for no match is
$\frac{\kappa}{N+\kappa}$.

\begin{equation}
P(z_{x,y}=1 | \alpha_x,\beta_{x,1}, \cdots, \beta_{x,N_x}) =
\frac{P(\alpha_x,\beta_{x,1}, \cdots, \beta_N
  |z_{x,y}=1)\frac{1}{N_x + \kappa}}{\sum_{j=1..N_x} P(\alpha_x,\beta_{x,1},
  \cdots,\beta_{x,N_x} |z_{x,j}=1) \frac{1}{N_x + \kappa} + P(\alpha_x,\beta_{x,1},
  \cdots,\beta_{x,N_x}|z_{x,1}=0, \cdots, z_{x,N_x}=0) \frac{\kappa}{N_x+\kappa}}
\label{quantityinterest}
\end{equation}

\noindent
This simplifies to

\begin{equation}
P(z_{x,y}=1 | \alpha_x,\beta_{x,1}, \cdots, \beta_{x,N_x}) =
\frac{P(\alpha_x,\beta_{x,1}, \cdots, \beta_{N_x}
  |z_{x,y}=1)}{\sum_{j=1..N_x} P(\alpha_x,\beta_{x,1},
  \cdots,\beta_{x,N_x} |z_{x,j}=1) + P(\alpha_x,\beta_{x,1},
  \cdots,\beta_{x,N_x}|z_{x,1}=0, \cdots, z_{x,N_x}=0) \kappa}
\label{quantityinterest2}
\end{equation}

\noindent
Substituting \ref{indepen} and \ref{indepen2} into \ref{quantityinterest} we get

\begin{eqnarray}
P(z_{x,y}=1 | \alpha_x,\beta_{x,1}, \cdots, \beta_{x,N_x}) = \nonumber \\
&\frac{
P(\alpha_x,\beta_{x,k}|z_{x,k}=1) \prod_{i=1..N_x,i \ne k} P(\alpha_x, \beta_{x,i}|z_{x,j}=0)
}{\sum_{j=1..N_x} 
P(\alpha_x,\beta_{x,j}|z_{x,j}=1) \prod_{i=1..N_x,
  i \ne j} P(\alpha_x, \beta_{x,i}|z_{x,i}=0)
 + \prod_{i=1..N_x}
P(\alpha_x, \beta_{x,j}|z_{x,j}=0)\kappa 
}
\end{eqnarray}

\noindent
Divide top and bottom by $\prod_{i=1..N_x} P(\alpha, \beta_i|z_{x,i}=0)$
\begin{equation}
P(z_{x,y}=1 | \alpha_x,\beta_{x,1}, \cdots, \beta_{x,N_x}) =
\frac{
\frac{P(\alpha_x,\beta_{x,y}|z_{x,y}=1)}{ P(\alpha_x, \beta_{x,y}|z_{x,y}=0)}
}{\sum_{j=1..N_x} 
\frac{P(\alpha_x,\beta_{x,j}|z_{x,j}=1)}{  P(\alpha_x, \beta_{x,j}|z_{x,j}=0)}
+ \kappa}
\end{equation}

\noindent
The above is the likelihood ratio result from \cite{sutherland}.  As argued in the body of this document another useful form for this equation is: 

\begin{equation}
P(z_{x,y}=1 | \alpha_x,\beta_{x,1}, \cdots, \beta_{x,N_x}) = \frac{
\frac{1}{P(z_{x,y}=1 | \alpha_x,\beta_{x,y})^{-1}-1}
}{
\sum_{j=1..N_x}\frac{1}{P(z_{i,j}=1 | \alpha_x,\beta_{x,j})^{-1}-1}
+ \frac{P(z_{i,j}=1)}{P(z_{i,j}=0)} \kappa
}
\end{equation}

\noindent
the advantage being that a neural network or Platt calibrated SVM
return a probability of the form $P(z_{x,y}=1 | \alpha_x,\beta_{x,y})$.

\end{document}